\newcommand{\CLASSINPUTtoptextmargin}{2.cm}
\newcommand{\CLASSINPUTbottomtextmargin}{2.5cm}

\documentclass[conference]{IEEEtran}
\IEEEoverridecommandlockouts
\usepackage{graphicx}   
\usepackage{caption}    
\usepackage{float}      
\usepackage{cite}
\usepackage{amsmath,amssymb,amsfonts}
\usepackage{algorithmic}
\usepackage{graphicx}
\usepackage{textcomp}
\usepackage[normalem]{ulem} 
\usepackage{xcolor}         
\usepackage{ 
algorithm, algorithmic}
\usepackage{cuted}
\usepackage{cases}

\allowdisplaybreaks
\usepackage{xcolor}
\def\BibTeX{{\rm B\kern-.05em{\sc i\kern-.025em b}\kern-.08em
    T\kern-.1667em\lower.7ex\hbox{E}\kern-.125emX}}
\begin{document}

\title{Multi-Active RIS-Assisted THz Cell-Free Systems: Spectral and Energy Efficiency Tradeoff} 
\author{\IEEEauthorblockN{Mario R. Camana,~\IEEEmembership{Member,~IEEE}, Zaid Abdullah,~\IEEEmembership{Member,~IEEE}, Carla E. Garcia,~\IEEEmembership{Member,~IEEE},\\Chandan Kumar Sheemar,~\IEEEmembership{Member,~IEEE}, Eva Lagunas,~\IEEEmembership{Senior Member,~IEEE}, Symeon Chatzinotas~\IEEEmembership{Fellow,~IEEE}}
\thanks{This work has been supported by the Smart Networks and Services Joint Undertaking (SNS JU) project TERRAMETA under the European Union’s Horizon Europe research and innovation programme under Grant Agreement No 101097101, including top-up funding by UK Research and Innovation (UKRI) under the UK government’s Horizon Europe funding guarantee. \textit{(Corresponding author: Mario R. Camana)}}
\thanks{Mario R. Camana, Zaid Abdullah, Carla E. Garcia, Chandan Kumar Sheemar, Eva Lagunas, and Symeon Chatzinotas are with the SnT department at the University of Luxembourg (email:\{mario.camana, zaid.abdullah, carla.garcia, chandankumar.sheemar, eva.lagunas, symeon.chatzinotas\}@uni.lu).}

}
\maketitle

\begin{abstract}
Reconfigurable intelligent surfaces (RISs) and cell-free massive multiple-input multiple-output (CF-mMIMO) are effective solutions for mitigating large path loss and inter-cell interference in terahertz (THz) systems. However, passive RISs are notably limited from double-fading attenuation, motivating the use of active RISs with power amplification to improve signal strength. In this paper, we investigate a multi-active RIS-aided wideband CF-mMIMO system for THz communications, considering low-resolution digital-to-analog converters (DACs) to optimize the spectral efficiency (SE)-energy efficiency (EE) tradeoff by adjusting precoding vectors and reflection coefficient response of the RISs, subject to power and minimum desirable per-user rate constraints. This leads to a highly complex and non-convex, multi-objective and fractional optimization problem. To solve it,  we propose a tailored quadratic transformation to manage the fractional form. This allows decomposition into two subproblems, which are iteratively solved via a successive convex approximation algorithm to optimize the precoding vectors and active RIS reflection coefficients until convergence. Numerical results demonstrate that the proposed active RIS-aided CF-mMIMO system effectively addresses propagation loss and limited scattering in THz communication, achieving superior EE and SE compared to conventional passive RIS across diverse scenarios. Furthermore, the integration of low-resolution DACs shows significant improvement in EE while preserving satisfactory communication performance.
\end{abstract}


\section{Introduction}

The advent of terahertz (THz) communications has opened new frontiers for ultra-high data rate transmission, while addressing the spectrum bottleneck and positioning it as a cornerstone for next-generation wireless networks. THz frequencies, typically ranging from 0.1 to 10 THz, offer vast bandwidths that can support data-intensive applications such as virtual reality and augmented reality, which demand microsecond latency and ultra-fast download speeds \cite{Ning2023}. However, there are major challenges related to signal propagation in THz communication, such as high spreading loss, limited scattering and severe molecular absorption loss, which pose significant challenges for reliable and efficient communication. 

Multiple-input-multiple-output (MIMO) systems are considered a practical solution to mitigate propagation loss by creating high-gain directional beams through beamforming technologies. However, network densification leads to increased inter-cell interference, necessitating a control unit to manage the network on a large geographical scale. In this context, cell-free massive MIMO (CF-mMIMO) is emerging as a promising technology to efficiently serve numerous users while effectively managing inter-cell interference. In CF-mMIMO, multiple access points (APs) are strategically distributed across the coverage area and connected through backhaul links to a central processing unit (CPU) to coordinate their cooperation \cite{Ngo2017}. This configuration is designed to minimize the distance between users and their nearest APs, effectively reducing path loss and enhancing signal reliability.

To further enhance network capacity, deploying additional distributed APs in a CF-mMIMO network entails significant costs and power consumption, which may render this approach impractical. Fortunately, the emerging technology of reconfigurable intelligent surfaces (RISs) offers an energy-efficient alternative to enhance spectral efficiency (SE), energy efficiency (EE), and network coverage. An RIS, composed of low-cost and energy-efficient reflective elements, has the ability to dynamically manipulate the wireless propagation environment by precisely controlling incident signals to achieve desired reflections. This capability is particularly suitable in dense urban environments where obstacles such as buildings and vehicles can obstruct direct communication paths.
In this context, passive RISs with programmable reflecting elements have garnered significant attention due to their low power consumption. Despite this advantage, signals reflected by passive RISs suffer significantly from the double-fading attenuation. To mitigate this issue, the concept of active RISs has emerged \cite{Long2021}, incorporating power amplification of incident signals to compensate for signal degradation and enhance transmission efficiency. In active RISs, each reflecting element incorporates active load impedance to amplify the incident signal, while still retaining the benefits of RISs without relying on complex and power-intensive RF chain components.
Unlike passive RIS, where performance gains are primarily obtained by increasing the number of reflecting elements, active RIS leverages power amplification to achieve a more direct and efficient signal boost. This approach allows for a more compact RIS design, making it particularly appealing for THz communications to overcome the propagation challenges.

Another major challenge for THz communications is the high power consumption of transceiver hardware, particularly due to the ultra-high sampling rates. Therefore, the use of high-resolution digital-to-analog converters (DACs) becomes problematic, as their power consumption scales exponentially with resolution and bandwidth. This makes full-resolution DACs energy-inefficient and impractical for THz systems. 
A promising energy-efficient approach is the use of low-resolution DACs, which has been recognized as an effective strategy to lower hardware complexity and power consumption \cite{Rib2018}. This approach effectively reduces power demands on fully digital transceivers while maintaining acceptable communication performance, achieving a favorable balance between SE and EE.

\subsection{Related Works and Motivation}
THz communications, along with the aforementioned enabling technologies, represent a promising avenue for the next-generation wireless systems. In the following, we present an overview of the recent contributions and advancements in this domain.

A novel sum-rate maximization framework for a passive RIS assisted single-cell THz system for an indoor scenario is presented in \cite{Ma2020}. The authors combine zero-forcing precoding for interference mitigation with a low-complexity RIS phase response optimization to finetune the transmission beams and maximize the performance. The problem of joint channel estimation and sum-rate maximization for a passive RIS-assisted THz system with a single multi-antenna user was investigated in \cite{Ma2020_2}. The authors consider a challenging scenario with no direct link and exploit reciprocity with time-division duplexing (TDD) to enable joint estimation and communication. 
A wideband RIS-assisted THz system serving multiple single-antenna users was investigated in \cite{Hao2021} to maximize the sum-rate by jointly optimizing hybrid beamforming vectors at the AP and the reflection response of the passive RIS. By considering array-of-subarrays at the THz transmitter and a passive sub-RIS architecture, the authors in \cite{Wang2022} proposed a sensor-based channel estimation method to infer the parameters of the channel paths, which are then used to design beamformers and the reflection matrix of each sub-RIS.

Further, the deployment of multiple passive RISs to assist a MIMO THz system was considered in \cite{Ning2021}, where a beam training strategy and a hybrid beamforming design were proposed to maximize the achievable rate. Under the same system configuration but by characterizing the THz channels using the Saleh-Valenzuela (SV) model and assuming only one RIS and one user, the authors in \cite{Ning2022} investigated the problem of weighted sum-rate maximization. Wideband THz communication with multiple passive RISs and single-antenna users was considered in \cite{Yan2023}, aiming to maximize the sum-rate by optimizing the reflection coefficient matrices and analog and digital beamforming matrices.

Regarding active RISs, few investigations have been conducted on their impact in THz systems. Namely, \cite{Hao2023} considered a THz integrated sensing and communication system assisted by an active RIS to serve a single user and maximize the illumination power of the target.  In \cite{Far2023}, the authors investigated the uplink sum-rate maximization in a unmanned aerial vehicle (UAV)-assisted THz system with an active RIS deployed at the UAV. The work in \cite{Le2025} considered a hybrid active-passive RIS for THz communications, taking hardware impairments into account.
However, none of the aforementioned contributions regarding active RIS in THz systems have yet investigated the performance when deployed with low-resolution DACs or CF systems. 
For microwave frequencies, active RISs have been investigated in \cite{Tuan2024} to maximize the sum-rate in multi-carrier communications, and in \cite{Phu2025} to maximize the EE in a rate-splitting multiple-access (RSMA) system with low-resolution DACs.


In the context of CF-mMIMO, recent advancements are presented in the following. 
 For sub-6 GHz frequencies, \cite{Zha2021} considered a multi-passive RIS-assisted wideband CF-mMIMO network to maximize the weighted sum-rate, and \cite{Fan2024} considered an uplink CF system aided by a single active RIS to maximize the sum-rate.
Considering multi-active RISs in CF networks to serve multiple single-antenna users for microwave frequencies, SE maximization was investigated in \cite{Tian2024}, EE fairness maximization in \cite{Wang2023}, and secure downlink transmission in \cite{Dong2025}. In millimeter wave, the max-min fairness problem in a CF-mMIMO with low-resolution analog-to-digital converters (ADCs)/DACs was investigated in \cite{Kim2022}. \cite{Lan2024} and \cite{Ma2023} considered a multi-passive RIS-aided CF-mMIMO system to maximize the sum-rate, and our previous work \cite{Cam2025} investigated a multi-passive RIS-aided RSMA CF-mMIMO system with low-resolution DACs to maximize the minimum rate of the users.
Research contributions on the CF-mMIMO systems assisted with active RIS are available in \cite{Fan2024,Tian2024,Wang2023,Dong2025}.
The scenario without RIS deployment in a THz system was considered in \cite{Jia2024} with the aim of selecting the nearest AP selection. In\cite{Zhu2024}, the authors considered the coupling effects between propagation angles and frequencies and proposed a novel transmit antenna selection scheme to maximize the transmission rate. In \cite{Abo2024}, prediction based novel hybrid beamforming design for THz systems is investigated. The extension considering passive RISs for a THz CF system was investigated in \cite{Su2024}, in which the authors introduced an additional time delay layer and proposed a hybrid beamforming design for sum-rate maximization for single-antenna multi-user scenario.

It is noteworthy that only a limited number of studies have explored the performance of CF-mMIMO THz communication systems. While recent contributions have investigated RIS-aided CF systems, they typically focus on passive RIS configurations and do not consider the practical aspect of low-resolution DACs. Furthermore, all the studies are limited to single antenna users. These components are particularly relevant for THz systems, where energy efficiency, hardware complexity, and signal strength are critical concerns. A comprehensive analysis of multi-active RIS-assisted CF systems incorporating low-resolution DACs and multi-antenna users still remains an open and important research direction for advancing THz communication technologies.


\subsection{Main Contributions}

This work presents the first comprehensive analysis of a multi-active RIS-assisted CF-mMIMO system with low-resolution DACs and its application to wideband THz communications with multi-antenna users. The main contributions of this paper are summarized as follows:
\begin{itemize}
\item We consider a multi-active RIS-assisted wideband THz CF-mMIMO system with multi-antenna users, where the APs employ low-resolution DACs to reduce hardware complexity and power consumption. Unlike traditional design criteria that focus solely on SE or EE independently, our aim is to investigate the optimal SE-EE tradeoff by jointly optimizing the precoding vectors and the reflection response of each active RIS, while considering realistic channel and power consumption models and constrained by a minimum individual desirable user rate, a maximum transmission power at the APs, and a maximum amplification power of the active RISs.
\item We propose a novel multi-objective problem formulation defined as a weighted-sum of SE-EE under the aforementioned constraints. The weights dictate the priority of the system, e.g. greater EE or SE. To solve this problem, we propose a quadratic transformation method to linearize the fractional form of the objective function. Consequently, the non-convex problem is decomposed into two subproblems to enable the design of a novel alternating optimization framework. Namely, we first propose a successive convex approximation (SCA)-based method to optimize the precoding vectors given the reflection coefficient matrices of the active RISs. Subsequently, with the optimized precoding vectors, we further optimize the reflection coefficient matrices of the active RISs based on first-order lower bound approximations. This procedure is repeated until convergence.
\item Extensive simulations under various system parameters are presented to validate that the proposed scheme achieves a better SE and EE tradeoff compared to conventional passive RIS and random phase selection benchmarks. The results also offer insightful analysis regarding the bit resolution at DACs, number of active RIS elements, number of APs, and number of scheduled users to achieve an optimal SE and EE tradeoff in active RIS-assisted THz CF systems.
\end{itemize}

 The remainder of this paper is organized as follows: Section II introduces the system model and formulates the SE-EE tradeoff problem. Section III presents the proposed solution to jointly optimize the precoding vectors and the reflecting coefficient matrices. Section IV illustrates the simulation results, and Section V draws the conclusions.

\textit{Notations}: Scalars, vectors and matrices are denoted with lowercase, lowercase bold, and uppercase bold letters, respectively. The operators $\mathbb{R}$ and $\mathbb{C}$ represent real and complex values, respectively.  $\Re(\cdot)$ indicate the real part of a complex number and ${\rm{Tr}}(\cdot)$ represents the trace operation. ${\cal C}{\cal N}\left( {\mu,{\sigma ^2}} \right)$  denotes a circularly symmetric complex Gaussian (CSCG) distribution with mean $\mu$ and variance ${\sigma ^2}$, $\left\|  \cdot  \right\|$ represents the Euclidean norm, and ${\left\|  \cdot  \right\|_F}$ denotes the Frobenius norm.
Superscripts ${\left\{  \cdot  \right\}^T}$, and ${\left\{  \cdot  \right\}^H}$ represent the transpose and Hermitian transpose, respectively. $ \otimes $ denotes the Kronecker product, and ${{\bf{I}}_N}$ is the $N \times N$ identity matrix. ${\rm{diag}}\left( {\bf{a}} \right)$ denotes the transformation of input vector ${\bf{a}}$ into a diagonal matrix and ${\rm{diag}}\left( {\bf{A}} \right)$ represents a diagonal matrix that preserves only the diagonal elements of matrix ${\bf{A}}$.

\section{System Model and Problem Formulation}
\begin{figure}[!t]
\centerline{\includegraphics[width=8 cm]{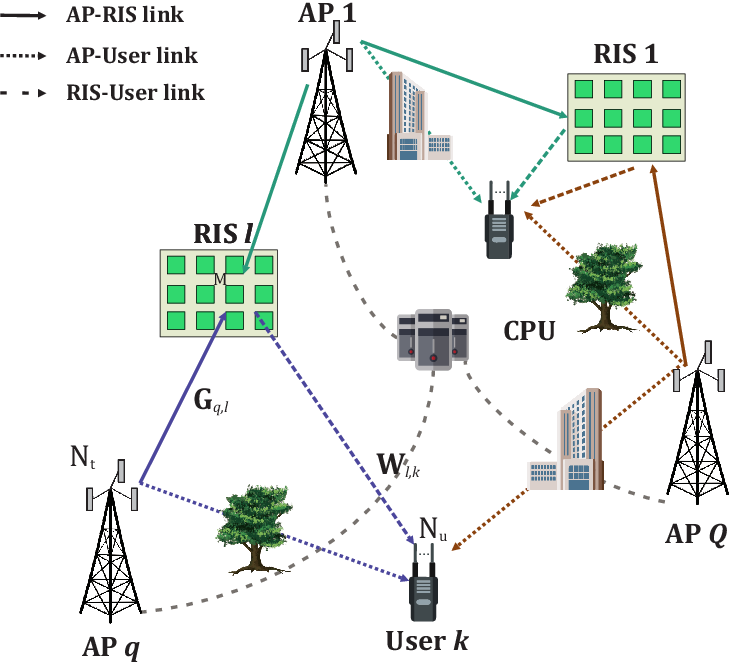}}
\caption{THz CF-mMIMO system assisted by multiple active RISs.}
\label{fig}
\end{figure}
We consider a multi-user wideband THz CF-mMIMO system assisted by multiple active RISs, as illustrated in Fig. 1. In this system, the $Q$ APs, each equipped with ${N_t}$ antennas, are assisted by $L \geq 1$ active RISs, each with $M$ elements, to support downlink multi-carrier transmission. We consider $B \geq 1$ available subcarriers to provide service to $K$ users, each equipped with ${N_u}$ antennas. The CPU is connected to all APs and is responsible for controlling and managing the network, with the active RISs being controlled by the CPU or the APs through control channels \cite{Zha2021}. Moreover, we assume an ideal feedback link between the AP and the RIS controller, and unlimited backhaul capacity from the APs to the CPU.
Given the high vulnerability of THz communications to obstructions, it is assumed that direct links between APs and users are blocked. We assume perfect channel state information (CSI)\footnote{This can be obtained with the CSI estimation methods reported in \cite{Liu2021}, \cite{Zhou2022}, \cite{Wan2023}.}. Furthermore, we incorporate low-resolution DACs at each AP to reduce the hardware cost and power consumption associated with the large antenna arrays required in THz communications \cite{Rib2018}, \cite{Zha2022}.  

Let ${{\bf{s}}_b} = {\left[ {{s_{b,1}},...,{s_{b,K}}} \right]^T} \in {\mathbb{C}^K}$ denotes the vector of data symbols transmitted over the $b$-th subcarrier intended for the $k$-th user, and $\mathbb{E}\left[ {{{\bf{s}}_b}{\bf{s}}_b^H} \right] = {{\bf{I}}_K}, \, \forall k$. At $q$-th AP, the symbol ${s_{b,k}}$ is precoded by the precoding vector ${{\bf{f}}_{q,k,b}} \in {\mathbb{C}^{{N_t} \times 1}}$, leading to a transmit signal before quantization on the $b$-th subcarrier given by
\begin{equation} 
        {{\bf{x}}_{q,b}} = \sum\limits_{k = 1}^K {{{\bf{f}}_{q,k,b}}} {s_{k,b}}.
\end{equation}

The quantization error of the low-resolution DACs at each AP are modeled using the additive quantization noise (AQN) \cite{Rib2018}, where the quantized transmitted signal can be represented by a linear function of the input signal ${{\bf{x}}_{q,b}}$ and uncorrelated noise. This allows to represent the quantized signal at the $q$-th AP on the $b$-th subcarrier as
\begin{equation} 
	{{\bf{\tilde x}}_{q,b}} = \sqrt {1 - {\alpha _q}} \sum\limits_{k = 1}^K {{{\bf{f}}_{q,k,b}}{s_{k,b}}}  + {{\bf{\tilde n}}_{q,b}},
\end{equation}
where ${{\bf{\tilde n}}_{q,b}}$ is the quantization noise uncorrelated with the input signal ${{\bf{x}}_{q,b}}$, and ${\alpha_q}$ represents the quantization distortion factor at the DACs of the $q$-th AP, where we assume that all the DACs at the AP have the same quantization bits, denoted as ${b_q^{DAC}}$. The value of ${\alpha_q}$ for ${b_q^{DAC}}>5$ can be approximated as ${\alpha _q} \approx {{\pi \sqrt 3 } \over 2}{2^{ - 2b_q^{DAC}}}$, while the values of ${\alpha_q}$ for up to five quantization bits are given by $\{0.3634,0.1175, 0.03454, 0.009497, 0.002499\}$ for ${b_q^{DAC}}=\{1,2,3,4,5\}$, respectively \cite{Xu2019}. Moreover, the covariance matrix of the quantization noise ${{\bf{\tilde n}}_{q,b}}$ is given by
\begin{equation} 
{{\boldsymbol{\Sigma }}_{{{{\bf{\tilde n}}}_{q,b}}}} = {\alpha _q}{\rm{diag}}\left( {\sum\limits_{k = 1}^K {{{\bf{f}}_{q,k,b}}} {\bf{f}}_{q,k,b}^H} \right).
\end{equation}
The channel response from the $q$-th AP to the $l$-th RIS, and from the $l$-th RIS to the $k$-th user on the $b$-th subcarrier are denoted as ${{\bf{G}}_{q,l,b}} \in {\mathbb{C}^{M \times {N_t}}}$, and ${{\bf{W}}_{l,k,b}} \in {\mathbb{C}^{M \times N_u}}$, respectively. 
The reflecting coefficient matrix of the $l$-th active RIS is represented as ${{\bf{\Theta }}_l} = {\rm{diag}}\left( \phi_{l,1},...,\phi_{l,M} \right)
  \in {\mathbb{C}^{M \times M}}$, with ${\phi _{l,m}} = {\beta _{l,m}}{e^{j{\theta _{l,m}}}}$, where  ${\beta _{l,m}} \in \left[ {0,{\beta _{m,\max }}} \right]$ and ${\theta _{l,m}} \in \left[ {0,2\pi } \right)$ represent the amplitude and the phase shift of the $m$-th element of $l$-th active RIS, with ${\beta _{m,\max }}$ denoting the maximum feasible value and in general ${\beta _{m,\max }}\ge 1$ \cite{Long2021}. The composite channel between the $q$-th AP and the $k$-th user, is expressed as ${\bf{J}}_{q,k,b}^H = \sum\limits_{l = 1}^L {{\bf{W}}_{l,k,b}^H{\bf{\Theta }}_l^H{{\bf{G}}_{q,l,b}}} $. Then, the received signal at the $k$-th user is given by
 \begin{align}
		{{\bf{y}}_{k,b}} =&\sum\limits_{q = 1}^Q {{\bf{J}}_{q,k,b}^H{{{\bf{\tilde x}}}_{q,b}}}  + \sum\limits_{l = 1}^L {{\bf{W}}_{l,k,b}^H{\bf{\Theta }}_l^H{{\bf{v}}_{l}}}  + {{\bf{n}}_{k,b}},\\
		=& \sum\limits_{q = 1}^Q {{\bf{J}}_{q,k,b}^H\left( {{\lambda _q}\sum\limits_{k' = 1}^K {{{\bf{f}}_{q,k',b}}} {s_{k',b}} + {{{\bf{\tilde n}}}_{q,b}}} \right)} + \nonumber \\
		&\sum\limits_{l = 1}^L {{\bf{W}}_{l,k,b}^H{\bf{\Theta }}_l^H{{\bf{v}}_l}}  + {{\bf{n}}_{k,b}},
\end{align}
where ${\lambda _q} = \sqrt {1 - {\alpha _q}} $, and ${{\bf{n}}_{k,b}} \sim \mathcal{CN}({\bf{0}}_{N_u},\sigma_k^2{\bf{I}}_{N_u})$ denotes the additive white Gaussian noise at the $k$-th user. 
The vector ${{\bf{v}}_l} \in {\mathbb{C}^{M \times 1}}$ represents the thermal noise at the $l$-th active RIS following ${\cal C}{\cal N}\left( {{{\bf{0}}_N},{\rm{ }}\sigma _{v,l}^2{{\bf{I}}_M}} \right)$ \cite{Long2021}, which is due to active hardware components. 

In the following, we introduce the compact notations to collect signal from all the $L$ RISs as
\begin{subequations}
    \begin{equation}
       {\bf{\Theta}} = \operatorname{blkdiag}({\bf{\Theta}}_1, {\bf{\Theta}}_2, \dots, {\bf{\Theta}}_L),
    \end{equation}
    \begin{equation}
        {\bf{v}} = {\left[ {{\bf{v}}_1^T,...,{\bf{v}}_L^T} \right]^T} \in {\mathbb{C}^{LM \times 1}},
    \end{equation}
    \begin{equation}
        {{\bf{W}}_{k,b}} = {\left[ {{\bf{W}}_{1,k,b}^T,....,{\bf{W}}_{L,k,b}^T} \right]^T} \in {\mathbb{C}^{LM \times {N_u}}},
    \end{equation}
\end{subequations}
where ${\bf{\Theta}}_l$, denoting the response of the $l$-th active RIS, represents the $l$-th block of the block diagonal matrix $\mathbf{\Theta}$. At the user, the receive filter ${{\boldsymbol{\omega }}_{k,b}} \in {\mathbb{C}^{1 \times {N_u}}}$ is applied to decode the stream at the $b$-th subcarrier. Then, the received signal at the $k$-th user after the receive filter on the $b$-th subcarrier can be expressed compactly as
\begin{equation}
    \begin{split}
        y_{k,b} = & \sum\limits_{q=1}^Q \lambda_q \boldsymbol{\omega}_{k,b} \mathbf{J}_{q,k,b}^H \mathbf{f}_{q,k,b} s_{k,b} \\
        &+ \sum\limits_{q=1}^Q \sum\limits_{\scriptstyle k' = 1 \hfill \atop 
					\scriptstyle k' \ne k \hfill}^K \lambda_q \boldsymbol{\omega}_{k,b} \mathbf{J}_{q,k,b}^H \mathbf{f}_{q,k',b} s_{k',b} \\
        & + \sum\limits_{q=1}^Q \boldsymbol{\omega}_{k,b} \mathbf{J}_{q,k,b}^H \tilde{\mathbf{n}}_{q,b} + \boldsymbol{\omega}_{k,b} \mathbf{W}_{k,b}^H \boldsymbol{\Theta}^H \mathbf{v} + \boldsymbol{\omega}_{k,b} \mathbf{n}_{k,b},
    \end{split}
\end{equation}
where the first term denotes the intended signal for the $k$-th user, the second term represents inter-user interference, and the third term accounts for distortion caused by low-resolution DACs at the APs. The fourth term corresponds to noise introduced by the active RISs, while the final term represents the receiver noise.

Furthermore, for the ease of notational convenience, we collect the response of $Q$ APs as follows
\begin{subequations}
    \begin{equation}
        {{\bf{J}}_{k,b}} = {\left[ {{\bf{J}}_{1,k,b}^T,....,{\bf{J}}_{Q,k,b}^T} \right]^T} \in {\mathbb{C}^{Q{N_t} \times {N_u}}},
    \end{equation}
    \begin{equation}
        {{\bf{f}}_{k,b}} = {\left[ {{\bf{f}}_{1,k,b}^T,....,{\bf{f}}_{Q,k,b}^T} \right]^T} \in {\mathbb{C}^{Q{N_t} \times 1}},
    \end{equation}
    \begin{equation}
        {{\boldsymbol{\chi }}_{k,b}} = {\rm{diag}}\left( {{\lambda _1},...,{\lambda _Q}} \right) \otimes {{\bf{I}}_{{N_t}}} \in {\mathbb{R}^{Q{N_t} \times Q{N_t}}},
    \end{equation}
\end{subequations}
which allows to write the signal-to-interference-plus-noise-ratio (SINR) at the $k$-th user compactly as \eqref{eq:SINR} at the top of the next page. 

\begin{figure*}[t]
	\centering
	\begin{equation} \label{eq:SINR}
		{\rm{SIN}}{{\rm{R}}_{k,b}} = {{{{\left| {{{\boldsymbol{\omega }}_{k,b}}{\bf{J}}_{k,b}^H{{\boldsymbol{\chi }}_{k,b}}{{\bf{f}}_{k,b}}} \right|}^2}} \over {\sum\limits_{\scriptstyle k' = 1 \hfill \atop 
					\scriptstyle k' \ne k \hfill} ^K {{{\left| {{{\boldsymbol{\omega }}_{k,b}}{\bf{J}}_{k,b}^H{{\boldsymbol{\chi }}_{k',b}}{{\bf{f}}_{k',b}}} \right|}^2}} +{\sum\limits_{q = 1}^Q {\left( {{{\boldsymbol{\omega }}_{k,b}}{\bf{J}}_{q,k,b}^H{{\boldsymbol{\Sigma }}_{{{{\bf{\tilde n}}}_{q,b}}}} {{\bf{J}}_{k,b}}{\boldsymbol{\omega }}_{k,b}^H} \right)} } + {{\left| {{{\boldsymbol{\omega }}_{k,b}}{\bf{W}}_{k,b}^H{{\boldsymbol{\Theta }}^H}{\bf{v}}} \right|}^2} + {{\left\| {{{\boldsymbol{\omega }}_{k,b}}} \right\|}^2}\sigma _{k,b}^2}}
	\end{equation} \hrulefill
\end{figure*}

\subsection{THz Channel Model With Multiple Active RISs}
A wideband THz system is considered with a central frequency $f_c$ and total bandwidth $B_W$. Then, the frequency sub-band corresponding to the $b$-th subcarrier is given by $f_b = f_c + \left( \frac{B_W}{B} \right)\left(b - 1 - \frac{B - 1}{2}\right), \quad b = 1, 2, \ldots, B$.
Note that THz waves experience significant free-space path loss due to their extremely high frequency and are readily absorbed by water molecules during airpropagation \cite{Ning2023}. Furthermore, in outdoor environment, due to high material absorption, lack of scattering and limited diffraction lead to negligible non-line-of-sight contributions. As a result, we consider only the LoS contributions in the channel response, where we consider that the APs and active RISs employ uniform planar arrays (UPAs) having $N_t = {N_{ty}} \times {N_{tz}}$ and $M = {M_{y}} \times {M_{z}}$, respectively, and the users employ uniform linear arrays (ULAs). Consequently, the channel matrices can be expressed as:
\begin{subequations} 
\begin{equation}
{{\bf{G}}_{q,l,b}} = {\alpha _{\bf{G}}}\left( {f_b},d_{q,l}^G \right){{\bf{a}}_{R,b}}\left( \gamma _{q,l}^{G,R},\eta _{q,l}^{G,R} \right){\bf{a}}_{T,b}^H\left( \gamma _{q,l}^{G,T},\eta _{q,l}^{G,T} \right)
\end{equation}

\begin{equation}
{{\bf{W}}_{l,k,b}} = {\alpha _{\bf{W}}}\left( {f_b},d_{l,k}^W \right){{\bf{a}}_{T,b}}\left( \gamma _{l,k}^{W,T},\eta _{l,k}^{W,T} \right){\bf{a}}_{ULA,b}^H\left( \zeta _{l,k}^{W,R} \right)
\end{equation}
\end{subequations}

where $d_{q,l}^G$ and $d_{l,k}^W$ denote the distance from the $q$-th AP to the $l$-th active RIS, and from the $l$-th active RIS to the $k$-th user, respectively.  
${\zeta _{l,k}^{W,R}}$ is the path angle of arrival in the RIS-user link, and ${\gamma _{q,l}^{G,R(T)},\eta _{q,l}^{G,R(T)}}$ represent the azimuth and elevation angle from the $b$-th AP to the $l$-th active RIS, where superscripts $(R)$ and $(T)$ indicate the arrival/reception and departure/transmission, respectively. The scalars ${\alpha _{\bf{G}}}$ and ${\alpha _{\bf{W}}}$ are defined as
${\alpha _{\bf{G}}}\left( {{f_b},d_{q,l}^G} \right) = \sqrt {{G_t}{N_t}M} \delta \left( {{f_b},d_{q,l}^G} \right)$ and ${\alpha _{\bf{W}}}\left( {{f_b},d_{l,k}^W} \right) = \sqrt {{G_r}M{N_u}} \delta \left( {{f_b},d_{l,k}^W} \right)$, where $\delta \left( {{f_b},d} \right)$ is the path loss, and $G_t$, $G_r$ denote the transmit and receive antenna gain.
${{\bf{a}}_{T\left( R \right),b}}\left( {\gamma ,\eta } \right)$, ${{\bf{a}}_{ULA,b}}\left( {\zeta  } \right)$ denote the antenna array response vector for UPA, having $N = {N_y} \times {N_z}$ elements, of the transmitter (receiver) and for ULA at the user, which are given by
\begin{subequations}
	\begin{align}
		{{\bf{a}}_{T\left( R \right),b}}\left( {\gamma ,\eta } \right) = &{1 \over {\sqrt N }}\left[ {1,...,{e^{j{{2\pi {d_A}{f_b}} \over c}\left( {{n_y}\sin \left( \gamma  \right)\sin \left( \eta  \right) + {n_z}\cos \left( \eta  \right)} \right)}},} \right. \nonumber\\
		&{\left. {...,{e^{j{{2\pi {d_A}{f_b}} \over c}\left( {\left( {{N_y} - 1} \right)\sin \left( \gamma  \right)\sin \left( \eta  \right) + \left( {{N_z} - 1} \right)\cos \left( \eta  \right)} \right)}}} \right]^T}\\
		{{\bf{a}}_{ULA,b}}\left( \zeta  \right) = &{1 \over {\sqrt N }}{\left[ {1,{e^{j{{2\pi {d_A}{f_b}} \over c}{\rm{sin}}\left( \zeta  \right)}},...,{e^{j{{2\pi {d_A}{f_b}\left( {N - 1} \right)} \over c}{\rm{sin}}\left( \zeta  \right)}}} \right]^T},
	\end{align}
\end{subequations} 
where $c$ is the speed of light, and ${d_A}$ represents the antenna spacing at the APs or element spacing in the active RIS set to be half of the wavelength of the central frequency $f_c$. The path loss $\delta \left( {{f_b},d} \right)$ for THz communications integrates the free spreading loss and the molecular absorption loss and can be expressed as \cite{Hao2021}, \cite{Ning2021}
\begin{equation}
	\delta \left( {{f_b},d} \right) = {c \over {4\pi {f_b}d}}{e^{ - {1 \over 2}\xi \left( {{f_b}} \right)d}},
\end{equation}
where $\xi \left( {{f_b}} \right)$ denotes the medium absorption factor.

\subsection{On the SE and EE Metrics}
\subsubsection{SE}

The achievable rate of the $k$-th user over the $b$-th subcarrier is given by:
\begin{equation}
    R_{k,b} = \log_2 \left( 1 + \mathrm{SINR}_{k,b} \right),
\end{equation}
where $\mathrm{SINR}_{k,b}$ denotes the signal-to-interference-plus-noise ratio for user $k$ on subcarrier $b$.

The total SE of the system is then expressed as the sum of the individual user-subcarrier rates:
\begin{equation}
    \rho_{\mathrm{SE}} = \sum_{k=1}^{K} \sum_{b=1}^{B} R_{k,b},
\end{equation}
where $K$ is the total number of users and $B$ is the total number of subcarriers.

\subsubsection{Power Consumption Model and EE}

The total power consumption of the system is modeled as following \cite{Rib2018},  \cite{Ngo2018}, \cite{Long2021}:
\begin{equation} 
	{P_{sys}} = \sum\limits_{q = 1}^Q {{P_{AP,q}}}  + \sum\limits_{l = 1}^L {{P_{RIS,l}}}  + {P_{st}},
\end{equation}
where ${{P_{AP,q}}}$ is the transmit power consumption at the $q$-th AP and can be modelled as
\begin{equation} 
	{{P_{AP,q}}}={\eta _A^{ - 1}{\rm{Tr}}\left( {\sum\limits_{k = 1}^K {\sum\limits_{b = 1}^B {{{\bf{f}}_{q,k,b}}{\bf{f}}_{q,k,b}^H} } } \right)},
\end{equation}
where ${\eta _A} \in \left( {0,1} \right]$ denotes the amplifier efficiency factor at the AP. 
${{P_{RIS,l}}}$ denotes the power reflected by the $l$-th active RIS, and can be modelled as 
\begin{equation}
\begin{split}
P_{RIS,l} = \eta_R^{-1} \Bigg( & \sum_{b=1}^B \sum_{q=1}^Q \sum_{k=1}^K \left\| \mathbf{\Theta}_l^H \mathbf{G}_{q,l,b} \lambda_q \mathbf{f}_{q,k,b} \right\|^2 \\
& + \sum_{b=1}^B \sum_{q=1}^Q \mathrm{Tr}\left( \mathbf{\Theta}_l^H \mathbf{G}_{q,l,b} \boldsymbol{\Sigma}_{\tilde{\mathbf{n}}_{q,b}} \mathbf{G}_{q,l,b}^H \mathbf{\Theta}_l \right) \\
& + \left\| \mathbf{\Theta}_l^H \right\|_F^2 \sigma_{v,l}^2 \Bigg),
\end{split}
\end{equation}

where ${\eta _R} \in \left( {0,1} \right]$ denotes the amplifier efficiency factor at the active RIS. ${P_{st}}$ represents the static power consumption of the system depending on the circuitry of the APs, active RISs, users and backhaul links, and is given by \cite{Rib2018},  \cite{Ngo2018}, \cite{Long2021}:
\begin{equation}
   \begin{split}
	{P_{st}} = &\sum\limits_{q = 1}^Q {{N_t}P_{c,q}^A}  + B\sum\limits_{q = 1}^Q {2{N_t}P_q^D\left( {{F_s},b_q^{DAC}} \right)} + \sum\limits_{k = 1}^K {P_{c,k}^U}  +\\
	&\sum\limits_{q = 1}^Q {P_q^B}  + L \times M \times \left( {{P_{RIS,c}} + {P_{RIS,DC}}} \right),
	\end{split}
\end{equation}
where $P_{c,q}^A$ denotes the power consumption of a single RF chain at the $q$-th AP, ${P_q^D\left( {{F_s},b_q^{DAC}} \right)}$ represents the power consumption of a DAC with a resolution of $b_q^{DAC}$ bits and sampling at a frequency of $F_s$, set to be the double of the subcarrier bandwidth, and can be modelled as \cite{Rib2018}
\begin{equation}
	{P_q^D\left( {{F_s},b_q^{DAC}} \right)} = 1.5 \times {10^{ - 5}} \cdot {2^{b_q^{DAC}}} + 9 \times {10^{ - 12}} \cdot b_q^{DAC} \cdot {F_s}.
\end{equation}
Then, ${P_{c,k}^U}$ denotes the static power at the $k$-th user and ${P_q^B}$ denotes the comsumption power due to the backhaul link from the $q$-th AP to the CPU. At each active RIS, ${P_{RIS,DC}}$ and ${{P_{RIS,c}}}$ represent the DC bias power consumption and the control and switch circuit power consumption at each reflecting element.

The EE of the considered system can be expressed as
\begin{equation}
    \rho_{\mathrm{EE}} = \frac{\rho_{\mathrm{SE}}}{P_{\mathrm{sys}}}.
\end{equation}

\subsection{Problem Formulation}
Remark that SE increases with transmit power consumption, reaching its maximum when all available transmit power is used. However, this approach is in conflict with the EE maximization, as EE seeks to balance SE and power consumption. Therefore, we focus on achieving the SE-EE tradeoff rather than prioritizing SE or EE individually. Moreover, to properly combine the EE and SE, which have different units, we include ${{P_{tot}}}$ defined as the total power budget of the system \cite{Niu2023}. For such a purpose, the optimization problem can be formally stated as

\begin{equation}
   f\big(\rho _{EE},\rho _{SE}\big) =\kappa{\rho _{EE}} +\left( {1 - \kappa } \right){{\rho _{SE}}} /{{P_{tot}}}
\end{equation}
where $0 \le \kappa  \le 1$ denotes the weighting factor, which dictates the priority of the EE metric. Note that in our objective, we consider dividing the SE efficiency by the constant $P_{tot}$, which is a fixed/constant power budget given by
\begin{equation}
   {P_{tot}} = \sum\limits_{q = 1}^Q {\eta _A^{ - 1}P_{q,\max }^A}  + \sum\limits_{l = 1}^L {\eta _R^{ - 1}P_{l,\max }^R}  + {P_{st}},
\end{equation}
The objective of doing so is to have a comparable performance between the SE and EE metrics. 

Our aim is to optimize the precoding vectors at the APs and the reflection coefficient matrices of the active RIS to achieve the best SE-EE tradeoff. For such a purpose, the SE-EE tradeoff optimization problem under the practical constraints can be formulated as 
 \begin{subequations}\label{OP1}
\begin{equation}
\hspace{-30mm}\max_{\{{\bf{f}}_{q,k,b}\}, \{{\phi}_{l,m}\}} \quad f\big(\rho_{EE}, \rho_{SE} \big)
\label{OP1_obj}
\end{equation}
\begin{equation}
R_{k,b} \ge R_k^{\text{th}}, \quad \forall k, \forall b
\label{C1_OP1}
\end{equation}
\begin{equation}
\hspace{26mm}\text{Tr} \left( \sum_{k=1}^K \sum_{b=1}^B {\bf{f}}_{q,k,b} {\bf{f}}_{q,k,b}^H \right) \le P_{q,\max}^A, \forall q,
\label{C2_OP1}
\end{equation}
\begin{equation}
\hspace{3mm}P_{RIS,l} \le P_{l,\max}^R, \quad \forall l
\label{C3_OP1}
\end{equation}
\begin{equation}
\hspace{6mm}|{\phi}_{l,m}| \le \beta_{\max}, \quad \forall m, \forall l
\label{C4_OP1}
\end{equation}
\end{subequations}

where $R_k^{th}$ is the minimum required rate at the $k$-th user in each subcarrier, $P_{q,\max }^A$ is the maximum available power at the $q$-th AP, $P_{l,\max }^R$ is the power budget for the power reflected at the $l$-th active RIS, and ${\beta _{\max }}$ is the maximum amplitude that the active-load reflecting element can deliver. Constraint \eqref{C1_OP1} ensures a minimum rate to be achieved by the users at each subcarrier, and \eqref{C3_OP1} represents the amplification power constraint at each active RIS to amplify the incident signal with active loads.  

It is noteworthy that the optimization problem is highly non-convex due to the fractional and logarithmic structure of the objective, which combines two conflicting metrics—spectral and energy efficiency—over complex-valued precoders and RIS coefficients. The SINR expressions are non-linear and coupled across users, while the constraints, including power budgets and modulus bounds on RIS elements, further restrict the feasible set in a non-convex manner. These factors make the problem analytically intractable and require iterative approximation methods to obtain locally optimal solutions.

\section{Problem Solution}
In this section, we present a novel algorithmic framework to address the problem \eqref{OP1}. Note that the EE in the objective function has a fractional form, and fractional programming methods can be adopted \cite{Shen2018}, where the quadratic transformation has shown superior performance compared with the traditional Dinkelbach’s transformation. 
In particular, the quadratic transformation decouples the numerators and the denominators of $\rho_{EE}$ through an auxiliary variable $\tau $, based on Theorem 1 in \cite{Shen2018}, to derive an equivalent non-fractional formulation given by $2\tau \sqrt{\rho_{SE}} -  \tau^2 P_{sys}$.

Consequently, we consider a tailored quadratic transformation for our case by introducing $\tau $, which allows us to transform the problem \eqref{OP1} as
 \begin{subequations} \label{OPSCA1}
\begin{align}
\max_{\{{\bf{f}}_{q,k,b}\}, \{{\phi}_{l,m}\}, \tau} \;\; & 2\kappa \tau \sqrt{\rho_{SE}} - \kappa \tau^2 P_{sys} + \left(1 - \kappa\right) \frac{\rho_{SE}}{P_{tot}} \label{OPSCA1_obj} \\
\text{s.t.} \quad &  \eqref{C1_OP1}-\eqref{C4_OP1} 
\end{align}
\end{subequations}
where the optimal value of $\tau $ is given by ${\tau ^*} = \sqrt {{\rho _{SE}}} /{P_{sys}}$. Subsequently, we consider decomposing the problem \eqref{OPSCA1} into two subproblems, and design a novel alternating optimization-based algorithm.  A sequential optimization framework is adopted wherein, initially, for fixed values of $\tau$ and the reflection coefficient matrices of the active RISs, an SCA-based algorithm is employed to optimize the precoding vectors. Subsequently, leveraging the optimized precoders, the reflection coefficient matrices of the active RISs are refined to further enhance system performance.

\subsection{Precoding Optimization}
In this section, we consider optimizing the digital beamformers, assuming the other variables fixed.
Let ${{\boldsymbol{\sigma }}_v} = {\rm{diag}}\left( {\sigma _{v,1}^2,...,\sigma _{v,L}^2} \right)$, and we reformulate the SINR of the $k$-th user as
\begin{equation}
	{\rm{SIN}}{{\rm{R}}_{k,b}}  = {{{{\left| {{{\boldsymbol{\omega }}_{k,b}}{\bf{J}}_{k,b}^H{{\bf{\chi }}_{k,b}}{{\bf{f}}_{k,b}}} \right|}^2}} \over {{\psi _{k,b}}\left( {{{\bf{f}}_{k,b}}} \right)}},
\end{equation}
where ${{\psi _{k,b}}\left( {{{\bf{f}}_{k,b}}} \right)}$ is presented in  \eqref{eq:SINRSCA1} at the top of the next page, where we use the property of ${{\bf{x}}^H}{\bf{x}} = {\left\| {\bf{x}} \right\|^2}$,  ${\bf{v}}{{\bf{v}}^H} = {{\boldsymbol{\sigma }}_v} \otimes {{\bf{I}}_M}$ and an equivalent expression for the covariance matrix of the quantization noise given by 
\begin{equation}
    {\alpha _q}{\rm{diag}}\left( {\sum\limits_{k = 1}^K {{{\bf{f}}_{q,k,b}}} {\bf{f}}_{q,k,b}^H} \right) = \sum\limits_{k = 1}^K {{\alpha _q}{\rm{diag}}\left( {{{\bf{f}}_{q,k,b}}} \right){\rm{diag}}\left( {{\bf{f}}_{q,k,b}^H} \right)}.
\end{equation}

\begin{figure*}[t]
	\centering
	\begin{equation} \label{eq:SINRSCA1}
    \begin{aligned}
        {\psi_{k,b}}\left( {\mathbf{f}_{k,b}} \right) = &\sum\limits_{k' = 1, k' \neq k}^K {\left| {\boldsymbol{\omega}_{k,b}}\mathbf{J}_{k,b}^H{\mathbf{\chi}_{k',b}}{\mathbf{f}_{k',b}} \right|^2} + \sum\limits_{q = 1}^Q {\sum\limits_{k' = 1}^K {\alpha_q{\left\| {\boldsymbol{\omega}_{k,b}}\mathbf{J}_{q,k,b}^H\mathrm{diag}\left( {\mathbf{f}_{q,k',b}} \right) \right\|^2}} } \\
        &+ {\boldsymbol{\omega}_{k,b}}\mathbf{W}_{k,b}^H{\mathbf{\Theta}^H}\left( {\boldsymbol{\sigma}_v} \otimes {\mathbf{I}_M} \right){\mathbf{\Theta}}{\mathbf{W}_{k,b}}{\boldsymbol{\omega}_{k,b}^H} + {\left\| {\boldsymbol{\omega}_{k,b}} \right\|^2}\sigma_{k,b}^2
    \end{aligned}
	\end{equation}  \hrulefill
\end{figure*}

Next, we introduce the auxiliary variable ${{\varsigma _{k,b}}}$ to represent the SINR of the $k$-th user on the $b$-th subcarrier and reformulate problem \eqref{OPSCA1}, given ${{\bf{\Theta }}_l}$ and $\tau$, as

\begin{subequations}\label{OPf1}
\begin{equation}
\begin{aligned}
\max_{\mathbf{f}_{q,k,b},\, \varsigma_{k,b}} \quad 
& 2\kappa \tau \sqrt{ \sum_{k=1}^K \sum_{b=1}^B \log_2(1 + \varsigma_{k,b}) }
- \kappa \tau^2 P_{\text{sys}} \\
& + \frac{1 - \kappa}{P_{\text{tot}}} \sum_{k=1}^K \sum_{b=1}^B \log_2(1 + \varsigma_{k,b})
\end{aligned}
\end{equation}
\begin{equation} \label{C1_OPf1}
\text{SINR}_{k,b} \geq \varsigma_{k,b}, \quad \forall k, \forall b
\end{equation}
\begin{equation}\label{C2_OPf1}
\log_2(1 + \varsigma_{k,b}) \geq R_k^{\text{th}}, \quad \forall k, \forall b 
\end{equation}
\begin{equation}
\text{\eqref{C2_OP1}, \eqref{C3_OP1}}
\end{equation}
\end{subequations}
To solve problem \eqref{OPf1}, we develop an SCA-based algorithm, where we first adopt the linear approximation method \cite{Mao2019} to approximate the non-convex terms of constraint \eqref{C1_OPf1} as
\begin{equation}
  \begin{split}
    {\varsigma _{k,b}} \le &{{2\Re\left( {{{\left( {{\bf{f}}_{k,b}^{\left( t \right)}} \right)}^H}{\bf{\chi }}_{k,b}^T{{\bf{J}}_{k,b}}{\boldsymbol{\omega }}_{k,b}^H{{\boldsymbol{\omega }}_{k,b}}{\bf{J}}_{k,b}^H{{\bf{\chi }}_{k,b}}{{\bf{f}}_{k,b}}} \right)} \over {\psi _{k,b}^{\left( t \right)}}} \\
	& - {{{{\left| {{{\boldsymbol{\omega }}_{k,b}}{\bf{J}}_{k,b}^H{{\bf{\chi }}_{k,b}}{\bf{f}}_{k,b}^{\left( t \right)}} \right|}^2}} \over {{{\left| {\psi _{k,b}^{\left( t \right)}} \right|}^2}}}{\psi _{k,b}},\,\forall k,\,\forall b \label{Rc_OPf1}
	\end{split}
\end{equation}
where $ \psi _{k,b}^{\left( t \right)}  $ represents $\psi _{k,b}$ calculated at the $t$-th iteration using the optimal ${{{\bf{f}}_{k,b}}}$ and given by
\begin{equation} \label{eq:SINRSCA2}
	\begin{split}
		\psi _{k,b}^{\left( t \right)} = &\sum\limits_{\scriptstyle k' = 1 \hfill \atop 
			\scriptstyle k' \ne k \hfill} ^K {{{\left| {{{\boldsymbol{\omega }}_{k,b}}{\bf{J}}_{k,b}^H{{\bf{\chi }}_{k',b}}{\bf{f}}_{k',b}^{\left( t \right)}} \right|}^2}}  \\
			& +\sum\limits_{q = 1}^Q {\sum\limits_{k' = 1}^K {{\alpha _q}{{\left\| {{{\bf{\omega }}_{k,b}}{\bf{J}}_{q,k,b}^H{\rm{diag}}\left( {{\bf{f}}_{q,k',b}^{\left( t \right)}} \right)} \right\|}^2}} }  \\
			& + {{\boldsymbol{\omega }}_{k,b}}{\bf{W}}_{k,b}^H{{\bf{\Theta }}^H}\left( {{{\boldsymbol{\sigma }}_v} \otimes {{\bf{I}}_M}} \right){\bf{\Theta }}{{\bf{W}}_{k,b}}{\boldsymbol{\omega }}_{k,b}^H  \hspace{-1mm}+ \hspace{-1mm}{\left\| {{{\boldsymbol{\omega }}_{k,b}}} \right\|^2}\sigma _{k,b}^2. 
		\end{split}
\end{equation} 
The reflected power constraint \eqref{C3_OP1} is reformulated, using ${{\bf{G}}_{l,b}} = \left[ {{{\bf{G}}_{1,l,b}},....,{{\bf{G}}_{Q,l,b}}} \right] \in {\mathbb{C}^{M \times Q{N_t}}}$, the property of $\left\| {\bf{A}} \right\|_F^2 = {\rm{Tr}}\left( {{\bf{A}}{{\bf{A}}^H}} \right) = {\rm{Tr}}\left( {{{\bf{A}}^H}{\bf{A}}} \right)$ and the equivalent expression for the covariance matrix of the quantization noise adopted in \eqref{eq:SINRSCA1}, as follows
\begin{equation}
	\begin{split}
		&\eta _R^{ - 1}\left( {\sum\limits_{k = 1}^K {\sum\limits_{b = 1}^B {{{\left\| {{\bf{\Theta }}_l^H{{\bf{G}}_{l,b}}{{\bf{\chi }}_{k,b}}{{\bf{f}}_{k,b}}} \right\|}^2}} }  + } \right.\\
		& \sum\limits_{b = 1}^B {\sum\limits_{q = 1}^Q {\sum\limits_{k = 1}^K {{\alpha _q}\left\| {{\bf{\Theta }}_l^H{{\bf{G}}_{q,l,b}}{\rm{diag}}\left( {{{\bf{f}}_{q,k,b}}} \right)} \right\|_F^2} } }  +\\
		&\left. {\left\| {{\bf{\Theta }}_l^H} \right\|_F^2\sigma _{v,l}^2} \right) \le P_{l,\max }^R,\,\forall l. \label{RPow_OPf1}
	\end{split}
\end{equation}
Therefore, given the initial feasible points ${{\bf{f}}_{k,b}^{\left( t \right)}}$, problem \eqref{OPf1} can beexpressed as
\begin{subequations} \label{OPf2}
\begin{equation}
\begin{aligned}
\max_{\{{\bf{f}}_{q,k,b}\}, \{{\varsigma}_{k,b}\}} \quad 
& 2\kappa \tau \sqrt{ \sum_{k=1}^K \sum_{b=1}^B \log_2(1 + \varsigma_{k,b}) } 
- \kappa \tau^2 P_{sys} \\
&+ \frac{(1 - \kappa) \sum_{k=1}^K \sum_{b=1}^B \log_2(1 + \varsigma_{k,b})}{P_{tot}}
\end{aligned}
\end{equation}
\begin{equation}\label{C2_OPf2}
\text{s.t. } \varsigma_{k,b} \ge 2^{R_k^{th}} - 1, \quad \forall k, \forall b 
\end{equation}
\begin{equation}
\eqref{Rc_OPf1}, \eqref{RPow_OPf1},\eqref{C2_OP1}
\end{equation}
\end{subequations}
where constraint \eqref{C2_OPf2} is a straightforward transformation of \eqref{C2_OPf1}. 
Problem \eqref{OPf2} is convex and can be solved using available software packages such as the CVX toolbox in MATLAB \cite{Gran2024}. Algorithm 1 formally summarizes the procedure to obtain the solution of problem \eqref{OP1} given ${{\bf{\Theta }}_l}$ and $\tau$.

%
\begin{algorithm}[!t]
    \caption{Proposed solution for problem \eqref{OP1}, given ${{\bf{\Theta}}_l}$ and $\tau$}
    \begin{algorithmic}[1]
        \STATE \textbf{Input:} Counter $t_1 = 0$, initial point ${{\bf{f}}_{k,b}^{(t_1)}}$, initial objective $f({{\bf{f}}_{k,b}^{(t_1)}}) = 0$, and tolerance $\varepsilon$.
        \STATE \textbf{repeat}
            \STATE \hspace{0.3cm} Solve problem \eqref{OPf2} with ${{\bf{f}}_{k,b}^{(t_1)}}$ and denote the solution\\ \hspace{0.3cm} as ${\bf{f}}_{k,b}^*$.
            \STATE \hspace{0.3cm} Update ${{\bf{f}}_{k,b}^{(t_1+1)}} \leftarrow {\bf{f}}_{k,b}^*$ and evaluate $f({{\bf{f}}_{k,b}^{(t_1+1)}})$.
            \STATE \hspace{0.3cm} Update $t_1 \leftarrow t_1 + 1$.
        \STATE \textbf{until} $\left| f({{\bf{f}}_{k,b}^{(t_1)}}) - f({{\bf{f}}_{k,b}^{(t_1-1)}}) \right| < \varepsilon$
        \STATE \textbf{Output:} ${\bf{f}}_{k,b}^*$
    \end{algorithmic}
    \label{alg1}
\end{algorithm}

\subsection{Optimizing  ${{\bf{\Theta }}_l}$ given ${{\bf{f}}_{q,k,b}}$ and $\tau$}
In this subsection, we aim to optimize the reflection coefficient matrices of the active RISs, given the precoding vectors of the APs. First, let ${{\boldsymbol{\Theta }}_l} = {\rm{diag}}\left( {\boldsymbol{\phi}} _l \right)$, where ${\boldsymbol{\phi}} _l = {\left[ {{\phi _{l,1}},...,{\phi _{l,M}}} \right]^T} \in {\mathbb{C}^{M \times 1}}$ with ${\phi _{l,m}} = {\beta _{l,m}}{e^{j{\theta _{l,m}}}}$. Then, by introducing ${\boldsymbol{\phi}}  = {\left[ {{\boldsymbol{\phi}}  _1^T,...,{\boldsymbol{\phi}}  _L^T} \right]^T} \in {\mathbb{C}^{LM \times 1}}$, ${\bf{\Theta }} = {\rm{diag}}({\boldsymbol{\phi}}  )$, ${{\bf{G}}_{q,b}} = {\left[ {{\bf{G}}_{q,1,b}^T,....,{\bf{G}}_{q,L,b}^T} \right]^T} \in {\mathbb{C}^{LM \times {N_t}}}$, ${{\bf{U}}_{q,k,b}} = {\rm{diag}}({{\boldsymbol{\omega }}_{k,b}}{\bf{W}}_{k,b}^H){{\bf{G}}_{q,b}} \in {\mathbb{C}^{LM \times {N_t}}}$, ${{\bf{U}}_{k,b}} = \left[ {{{\bf{U}}_{1,k,b}},...,{{\bf{U}}_{Q,k,b}}} \right] \in {\mathbb{C}^{LM \times Q{N_t}}}$, ${{\bf{\Sigma }}_{{{{\bf{\tilde n}}}_b}}} = {\rm{diag}}\left( {{{\bf{\Sigma }}_{{{{\bf{\tilde n}}}_{1,b}}}},...,{{\bf{\Sigma }}_{{{{\bf{\tilde n}}}_{Q,b}}}}} \right)$,
and the equivalent expressions of $   \boldsymbol{\omega}_{k,b} \mathbf{J}_{k,b}^H = \boldsymbol{\phi}^H \mathbf{U}_{k,b}, \quad \mathbf{v}\mathbf{v}^H = \boldsymbol{\sigma}_v \otimes \mathbf{I}_M, \quad \boldsymbol{\omega}_{k,b} \mathbf{W}_{k,b}^H \boldsymbol{\Theta}^H = \boldsymbol{\phi}^H \mathrm{diag}(\boldsymbol{\omega}_{k,b} \mathbf{W}_{k,b}^H)$, the SINR of the $k$-th user on the $b$-th subcarrier can be reformulated as \eqref{SINRThet1} at the top of the next page.

\begin{figure*}[t]
    \centering
    \begin{equation} \label{SINRThet1}
    \mathrm{SINR}_{k,b} = 
    \frac{
        \left| \boldsymbol{\phi}^H \mathbf{U}_{k,b} \boldsymbol{\chi}_{k,b} \mathbf{f}_{k,b} \right|^2
    }{
        \displaystyle
        \sum_{\substack{k' = 1 \\ k' \neq k}}^K 
            \left| \boldsymbol{\phi}^H \mathbf{U}_{k,b} \boldsymbol{\chi}_{k',b} \mathbf{f}_{k',b} \right|^2
        + \boldsymbol{\phi}^H \mathbf{U}_{k,b} \boldsymbol{\Sigma}_{\tilde{\mathbf{n}}_b} \mathbf{U}_{k,b}^H \boldsymbol{\phi}
        + \boldsymbol{\phi}^H \mathrm{diag}\left( \boldsymbol{\omega}_{k,b} \mathbf{W}_{k,b}^H \right)
            \left( \boldsymbol{\sigma}_v \otimes \mathbf{I}_M \right)
            \mathrm{diag}\left( \mathbf{W}_{k,b} \boldsymbol{\omega}_{k,b}^H \right) \boldsymbol{\phi}
        + \left\| \boldsymbol{\omega}_{k,b} \right\|^2 \sigma_{k,b}^2
    }
    \end{equation}
    \hrulefill
\end{figure*}

Next, we introduce the auxiliary variables ${{\delta _{k,b}}}$ to denote the SINR of the $k$-th user and reformulate problem \eqref{OPSCA1}, given ${{\bf{f}}_{q,k,b}}$  and $\tau$, as follows:
\begin{subequations} \label{OPT1}
	\begin{align}
		\mathop {{\rm{max}}}\limits_{{\phi _{l,m}},{\delta _{k,b}}}    \,\,\, 
		&2\kappa \tau \sqrt {\sum\limits_{k = 1}^K {\sum\limits_{b = 1}^B {{{\log }_2}\left( {1 + {\delta _{k,b}}} \right)} } }  - \kappa {\tau ^2}{P_{sys}} + \nonumber \\
		&\left( {1 - \kappa } \right){{\sum\limits_{k = 1}^K {\sum\limits_{b = 1}^B {{{\log }_2}\left( {1 + {\delta _{k,b}}} \right)} } } \over {{P_{tot}}}}  \\
		\text{s.t. }  & {\rm{SIN}}{{\rm{R}}_{k,b}} \ge {\delta _{k,b}},\,\forall k,\,\forall b \label{C1_OPT1}\\
		&{\log _2}\left( {1 + {\delta _{k,b}}} \right) \ge R_k^{th},\,\forall k,\,\forall b\\
		&\left| {{\phi _{l,m}}} \right| \le {\beta _{\max }},\,\forall l,\,\forall m  \label{C3_OPT1}\\
		&\text{\eqref{C3_OP1}}.
	\end{align}
\end{subequations}

First, the non-convex constraint \eqref{C1_OPT1} is linearly approximated and expressed as

\begin{equation}
\begin{split}
\delta_{k,b} \leq 
    & \frac{2\, \Re\left( \mathbf{f}_{k,b}^H \boldsymbol{\chi}_{k,b}^\mathsf{T} \mathbf{U}_{k,b}^H \boldsymbol{\phi}^{(t)} \boldsymbol{\phi}^{H} \mathbf{U}_{k,b} \boldsymbol{\chi}_{k,b} \mathbf{f}_{k,b} \right)}{\psi_{\phi, k, b}^{(t)}} \\
    & - \frac{\left| \left( \boldsymbol{\phi}^{(t)} \right)^H \mathbf{U}_{k,b} \boldsymbol{\chi}_{k,b} \mathbf{f}_{k,b} \right|^2}{\left| \psi_{\phi, k, b}^{(t)} \right|^2} \psi_{\phi, k, b}, \quad \forall k,\, \forall b
\end{split}
\label{Rc_OPT1}
\end{equation}

where ${{\psi _{\phi ,k,b}}}$ is the denominator term from \eqref{SINRThet1}, and $\psi _{\phi ,k,b}^{\left( t \right)}$ is given in \eqref{Thet2} at the top of the next page.

\begin{figure*}[t]
	\centering
	\begin{equation} \label{Thet2}
		\begin{split}
		\psi _{\phi ,k,b}^{\left( t \right)} = \sum\limits_{\scriptstyle k' = 1 \hfill \atop 
			\scriptstyle k' \ne k \hfill} ^K {{{\left| {{{\left( {{\phi ^{\left( t \right)}}} \right)}^H}{{\bf{U}}_{k,b}}{{\bf{\chi }}_{k',b}}{{\bf{f}}_{k',b}}} \right|}^2}}   &+ {\left( {{\phi ^{\left( t \right)}}} \right)^H}{{\bf{U}}_{k,b}}\left( {{{\bf{\sigma }}_e} \otimes {{\bf{I}}_{{N_t}}}} \right){\bf{U}}_{k,b}^H{\phi ^{\left( t \right)}} \\&+ {\left( {{\phi ^{\left( t \right)}}} \right)^H}{\rm{diag}}({{\boldsymbol{\omega }}_{k,b}}{\bf{W}}_{k,b}^H)\left( {{{\bf{\sigma }}_v} \otimes {{\bf{I}}_M}} \right){\rm{diag}}({{\bf{W}}_{k,b}}{\boldsymbol{\omega }}_{k,b}^H){\phi ^{\left( t \right)}}+ {\left\| {{{\boldsymbol{\omega }}_{k,b}}} \right\|^2}\sigma _{k,b}^2.
		\end{split}
	\end{equation} \hrulefill
\end{figure*}


Next, we reformulate the reflected power constraint \eqref{C3_OP1} of the active RIS as following:
	\begin{equation}
	\begin{split}
		\eta _R^{ - 1}\left( {{{\left\| {{\phi _l}} \right\|}^2}\sigma _{v.l}^2 + } \right.\sum\limits_{b = 1}^B {{\rm{Tr}}\left( {{\rm{diag}}{{\left( {{\phi _l}} \right)}^H}{{\bf{G}}_{l,b}}{{\bf{\Sigma }}_{{{{\bf{\tilde n}}}_b}}}{\bf{G}}_{l,b}^H{\rm{diag}}\left( {{\phi _l}} \right)} \right)}  +  \\
		\left. {\sum\limits_{k = 1}^K {\sum\limits_{b = 1}^B {{{\left\| {{\rm{diag}}\left( {{{\bf{G}}_{l,b}}{{\bf{\chi }}_{k,b}}{{\bf{f}}_{k,b}}} \right){\phi _l}} \right\|}^2}} } } \right) \le P_{l,\max }^R,\,\forall l. \label{RPow_OPT1}
	\end{split}
\end{equation}

Then, given the initial feasible points ${{\phi ^{\left( t \right)}}}$, problem \eqref{OPT1} is expressed as
\begin{subequations} \label{OPT2}
	\begin{align}
		\mathop {{\rm{max}}}\limits_{{\phi _{l,m}},{\delta _{k,b}}}    \,\,\, 
		&2\kappa \tau \sqrt {\sum\limits_{k = 1}^K {\sum\limits_{b = 1}^B {{{\log }_2}\left( {1 + {\delta _{k,b}}} \right)} } }  - \kappa {\tau ^2}{P_{sys}} + \nonumber \\
		&\left( {1 - \kappa } \right){{\sum\limits_{k = 1}^K {\sum\limits_{b = 1}^B {{{\log }_2}\left( {1 + {\delta _{k,b}}} \right)} } } \over {{P_{tot}}}} \\
		\text{s.t. }  & {\delta _{k,b}} \ge {2^{R_k^{th}}} - 1,\,\forall k,\,\forall b\\
		&\text{\eqref{C3_OPT1}, \eqref{Rc_OPT1}, and \eqref{RPow_OPT1}}.
	\end{align}
\end{subequations}

Problem \eqref{OPT2} is convex and the CVX toolbox in MATLAB \cite{Gran2024} can be used to obtain the solution. Algorithm 2 outlines the procedure for solving problem \eqref{OP1} given ${{\bf{f}}_{q,k,b}}$ and $\tau$.
 Finally, Algorithms 1 and 2 are optimized iteratively to obtain the precoding vectors, and reflection coefficient matrices until convergence is achieved, as presented in Algorithm 3. The variable  
 $\tau $ and the receive filter ${{\boldsymbol{\omega }}_{k,b}}$ at the $k$-th user on the $b$-th subcarrier are updated at each iteration, with ${{\boldsymbol{\omega }}_{k,b}}$ given by ${{\boldsymbol{\omega }}_{k,b}} = {{\boldsymbol{\tilde \omega }}_{k,b}}/\left\| {{{{\boldsymbol{\tilde \omega }}}_{k,b}}} \right\|$, where ${{\boldsymbol{\tilde \omega }}_{k,b}}$ is obtained using the minimum mean-square error (MMSE) solution as follows \cite{Vuci2009}:
 \begin{equation} \label{Wxu}
\begin{split}
\boldsymbol{\tilde{\omega}}_{k,b} =\, & 
\mathbf{f}_{k,b}^H \mathbf{J}_{k,b} \Bigg( 
    \sum_{k' = 1}^K \mathbf{J}_{k,b}^H \boldsymbol{\chi}_{k',b} \mathbf{f}_{k',b} \left( \mathbf{J}_{k,b}^H \boldsymbol{\chi}_{k',b} \mathbf{f}_{k',b} \right)^H + \sigma_{k,b}^2 \mathbf{I}_{N_u}\\
    & + \mathbf{J}_{k,b}^H \boldsymbol{\Sigma}_{\tilde{\mathbf{n}}_b} \mathbf{J}_{k,b} 
    + \mathbf{W}_{k,b}^H \boldsymbol{\Theta}^H \left( \boldsymbol{\sigma}_v \otimes \mathbf{I}_M \right) \boldsymbol{\Theta} \mathbf{W}_{k,b} 
\Bigg)^{-1}
\end{split}
\end{equation}

\begin{algorithm}[!t]
\caption{Proposed solution for problem \eqref{OP1}, given ${{\bf{f}}_{q,k,b}}$ and $\tau$}
\begin{algorithmic}[1]
	\STATE \textbf{inputs:}  Counter $t_2$=0, initial points ${{\boldsymbol{\phi}} ^{\left( t_2 \right)}}$, initial objective function $f( {{\boldsymbol{\phi}} ^{\left( t_2 \right)}}) = 0$, and tolerance $\varepsilon  $.
	\STATE \textbf{repeat}  
	\STATE \hspace{0.3cm} Solve problem \eqref{OPT2} with ${{\boldsymbol{\phi}} ^{\left( t_2 \right)}}$ and denote the \\
	\hspace{0.3cm} solution as  ${{\boldsymbol{\phi}} ^*}$.\\
	\STATE \hspace{0.3cm} Update ${{\boldsymbol{\phi}} ^{\left( t_2+1 \right)}} \leftarrow {{\boldsymbol{\phi}} ^*}$ and evaluate $f( {{\boldsymbol{\phi}} ^{\left( t_2+1\right)}})$.
	\STATE \hspace{0.3cm} Update $t_2 \leftarrow t_2+1$.
	\STATE \textbf{until} $f( {{\boldsymbol{\phi}} ^{\left( t_2+1 \right)}}) -f( {{\boldsymbol{\phi}} ^{\left( t_2 \right)}}) < \varepsilon $
	\STATE \textbf{outputs:} ${{\boldsymbol{\phi}} ^*}$.
\end{algorithmic}
\label{alg1}
\end{algorithm}



    
\begin{algorithm}[!t]
\caption{Proposed joint AP precoding and active RIS design to solve problem \eqref{OP1}}
\begin{algorithmic}[1]
    \STATE \textbf{Input:} Counter $t_3 = 0$, initial points ${\boldsymbol{\phi}}^{(t_3)}$, ${\mathbf{f}}_{k,b}^{(t_3)}$, initial objective $f\left({\boldsymbol{\phi}}^{(t_3)}, {\mathbf{f}}_{k,b}^{(t_3)}\right) = 0$, and tolerance $\varsigma$.
    \REPEAT
        \STATE \hspace{0.3cm} Update ${\boldsymbol{\tilde{\omega}}}_{k,b}$, $\forall k$ by \eqref{Wxu}.
        \STATE \hspace{0.3cm} Update $\tau^* = \sqrt{\rho_{SE}} / P_{sys}$.
        \STATE \hspace{0.3cm} Update ${\mathbf{f}}_{k,b}^*$ by solving problem \eqref{OPf2} with\\
        \hspace{0.3cm} Algorithm~1.
        \STATE \hspace{0.3cm} Update ${\boldsymbol{\phi}}^*$ by solving problem \eqref{OPT2} with \\
        \hspace{0.3cm} Algorithm~2.
        \STATE \hspace{0.3cm} Evaluate $f\left({\boldsymbol{\phi}}^{(t_3+1)}, {\mathbf{f}}_{k,b}^{(t_3+1)}\right)$.
        \STATE \hspace{0.3cm} Update $t_3 \gets t_3 + 1$.
    \UNTIL{$\left| f\left({\boldsymbol{\phi}}^{(t_3)}, {\mathbf{f}}_{k,b}^{(t_3)}\right) - f\left({\boldsymbol{\phi}}^{(t_3-1)}, {\mathbf{f}}_{k,b}^{(t_3-1)}\right) \right| < \varsigma$}
    \STATE \textbf{Output:} ${\boldsymbol{\phi}}^*, {\mathbf{f}}_{k,b}^*$
\end{algorithmic}
\label{alg1}
\end{algorithm}

\subsection{Convergence}
The convergence of the proposed alternating optimization algorithm for solving problem \eqref{OP1} can be established by verifying that the objective function is non-decreasing and bounded above. 

Specifically, at each iteration, the algorithm alternately updates the AP precoding vectors \( \{{\bf{f}}_{q,k,b}\} \) and the active RIS coefficients \( \{{\phi}_{l,m}\} \) by solving the subproblems \eqref{OPf2} and \eqref{OPT2}, respectively. Given the RIS coefficients and auxiliary variable \( \tau \), the precoding vectors are optimized by solving the convex approximation of problem \eqref{OPf2} using SCA, where the non-convex SINR constraints are replaced by convex upper bounds as shown in \eqref{Rc_OPf1}. Similarly, for fixed precoding vectors, the active RIS coefficients are optimized by solving the convexified problem \eqref{OPT2}, after approximating the original non-convex SINR expressions via linearization, as described in \eqref{Rc_OPT1}. At each subproblem update, the objective function \( f(\rho_{\mathrm{EE}}, \rho_{\mathrm{SE}}) \), defined in \eqref{OP1_obj}, either increases or remains constant since convex surrogate functions are tight and optimally solved at each step. Moreover, the auxiliary variable \( \tau \) is updated using the closed-form expression \( \tau^{*} = \sqrt{\rho_{\mathrm{SE}}}/P_{\mathrm{sys}} \), as derived by optimizing \eqref{OPSCA1_obj}, and the receive filters \( \boldsymbol{\omega}_{k,b} \) are updated according to the MMSE solution in \eqref{Wxu} \cite{Vuci2009}, both of which guarantee non-decreasing improvement of the objective value. 

Note that the achievable spectral efficiency \( \rho_{\mathrm{SE}} \) and energy efficiency \( \rho_{\mathrm{EE}} \) are non-negative and upper-bounded due to the limited power budget and finite channel gains. Consequently, the sequence of objective values generated by the proposed algorithm is monotonically non-decreasing and bounded, implying convergence. Although the original problem \eqref{OP1} is non-convex, global optimality cannot be guaranteed; however, the algorithm converges to a stationary point satisfying the Karush-Kuhn-Tucker (KKT) conditions of the tightly approximated subproblems. Thus, the convergence of the proposed joint AP precoding and active RIS optimization framework is formally established.

\subsection{Computational Complexity Analysis}
In the following, we aim to evalaute the per-iteration computational complexity of the proposed joint optimization framework.

Note that the per iteration complexity for solving the joint problem with Algorithm $3$, depends on the two inner loops implemented with Algorithms 1 and 2. Algorithm 1 solves subproblem \eqref{OPf2}, which is composed of $Q \times K \times B$ variables of size $N_t$ and $K \times B$ real variables. The computational complexity to solve \eqref{OPf2} is given by ${\psi _{\bf{f}}}=\mathcal{O}\left( T_1{{{\left( {KB + QKBN_t + L + Q} \right)}^{3.5}}   } \right)$ \cite{Meha2015}, \cite{Cam2022}, where $T_1$ is the total number of iterations required. Subsequently, Algorithm 2 solves subproblem \eqref{OPT2} and involves $L$ variables of size $M$ and $K \times B$ real variables.  The resultant computational complexity to solve \eqref{OPT2} can be determined by evaluating the number of mathematical operations required and is given by ${\psi _\phi }=\mathcal{O}\left( T_2{{{\left( {KB + LM + L} \right)}^{3.5}}} \right)$, where $T_2$ denotes the number of iterations required for Algorithm 2 to achieve convergence. Consequently, the per iteration computational complexity of Algorithm 3 is given by $\mathcal{O}\left({\psi _{\bf{f}}} + {\psi _\phi } \right)$.



\section{Simulation Results}
In this section, we present the numerical simulations under various system parameters to evaluate the performance of the proposed scheme.
We considered the THz channel and power consumption models detailed in Section II, operating at a frequency of 0.14 THz and having a total bandwidth of $B_W =  5$ GHz, with $B=4$ subcarriers, each having 1.25 GHz per subcarrier, as in \cite{Hao2021}.
The noise power density is assumed to be -174 dBm/Hz \cite{Hao2021},  and the transmit and receive antenna gain (in dBi) are set to $G_t=G_r=4 + 10{\rm{lo}}{{\rm{g}}_{10}}\left( {\sqrt {{N_{t/u}}} } \right)$ \cite{Ning2021}. The azimuth and elevation path angles are randomly selected from $U\left[-\pi/2,\pi/2\right]$. 
The $(x, y, z)$-coordinates of the $q$-th APs are given by $\left(12\left(q-1\right) / \left(Q-1\right),-4,6)\right)$ m, while the locations of the two active RIS are (5,3,6) m, and (8,3,6) m, respectively. The users are located in a square area of $3 \times3$ m, given by ($d_U$+$d_x$, $d_y$, 1.65) m, where $d_x$ and $d_y$ are randomly selected within $\left[0,3\right]$.
The default simulation parameters are presented in Table \ref{tablePara} unless otherwise indicated.

\begin{table}[]
	\caption{Simulation parameters}
	\label{tablePara}
	\begin{tabular}{|l|l|l|l|}
		\hline
		\textbf{Parameter} & \textbf{Value} & \textbf{Parameter} & \textbf{Value} \\ \hline \hline
		$f_c$ &  0.14 THz              &                   $Q$ & 3               \\ \hline
		$K$ &  4               &                    $N_u$ & 4               \\ \hline
		$L$ & 2               &                    $M$ & 64               \\ \hline
		$R_k^{th}$ & 0.1 (bit/s/Hz)                     &                    $d_U$  & 5 m                \\ \hline
		$P_{l,\max }^R$  \cite{Long2021} & 20 dBm                & ${\beta _{\max }^2}$ \cite{Long2021} & 20 dB   \\ \hline                
		${P_{c,k}^U}$ \cite{Wan2023}& 100mW &   ${P_q^B}$ \cite{Ngo2018} & 825mW   \\ \hline
		${P_{RIS,DC}}$ \cite{Long2021} &  -5dBm        & ${P_{RIS,DC}}$ \cite{Long2021}  &  -10 dBm \\ \hline
		${\eta _A}$ \cite{Niu2023} &  0.9       &${\eta _R}$  \cite{Long2021} & 0.8  \\ \hline
		$\xi \left(0.14 \text{THz}\right) $ \cite{Ning2021}& $6 \times 10^-5/m$   &  $P_{c,q}^A$ \cite{Rib2018}  &31.6 mW   \\ \hline
		
	\end{tabular}
\end{table}

The proposed scheme is denoted as ``ARIS", where the value of the weight parameter $ \kappa=1$ defines the objective function as EE, $ \kappa=0$ determines the optimization of the SE, and the trade-off between SE and EE is defined with the range $0 \le \kappa  \le 1$.
As a comparative scheme, we consider the passive RIS (PRIS), which does not account for thermal noise at the RIS and does not consume transmit power or DC bias power, i.e., $\sigma_{v,l}^{PRIS} = 0$, $P_{RIS,l}^{PRIS} = 0$, and $P_{RIS,DC}^{PRIS} = 0$.
Moreover, we consider the random phase shift selection (RND-ARIS), where the phase shift values of the elements in the active RIS are randomly selected within $\left[0, 2\pi\right)$, and the amplitude is set to $\beta_{\text{max}}$. Additionally, the baseline methods of PRIS and RND-RIS do not incorporate the minimum rate constraint to guarantee feasibility in all considered scenarios.

\begin{figure}[!t]
\centering
\includegraphics[width=9cm,height=7cm]{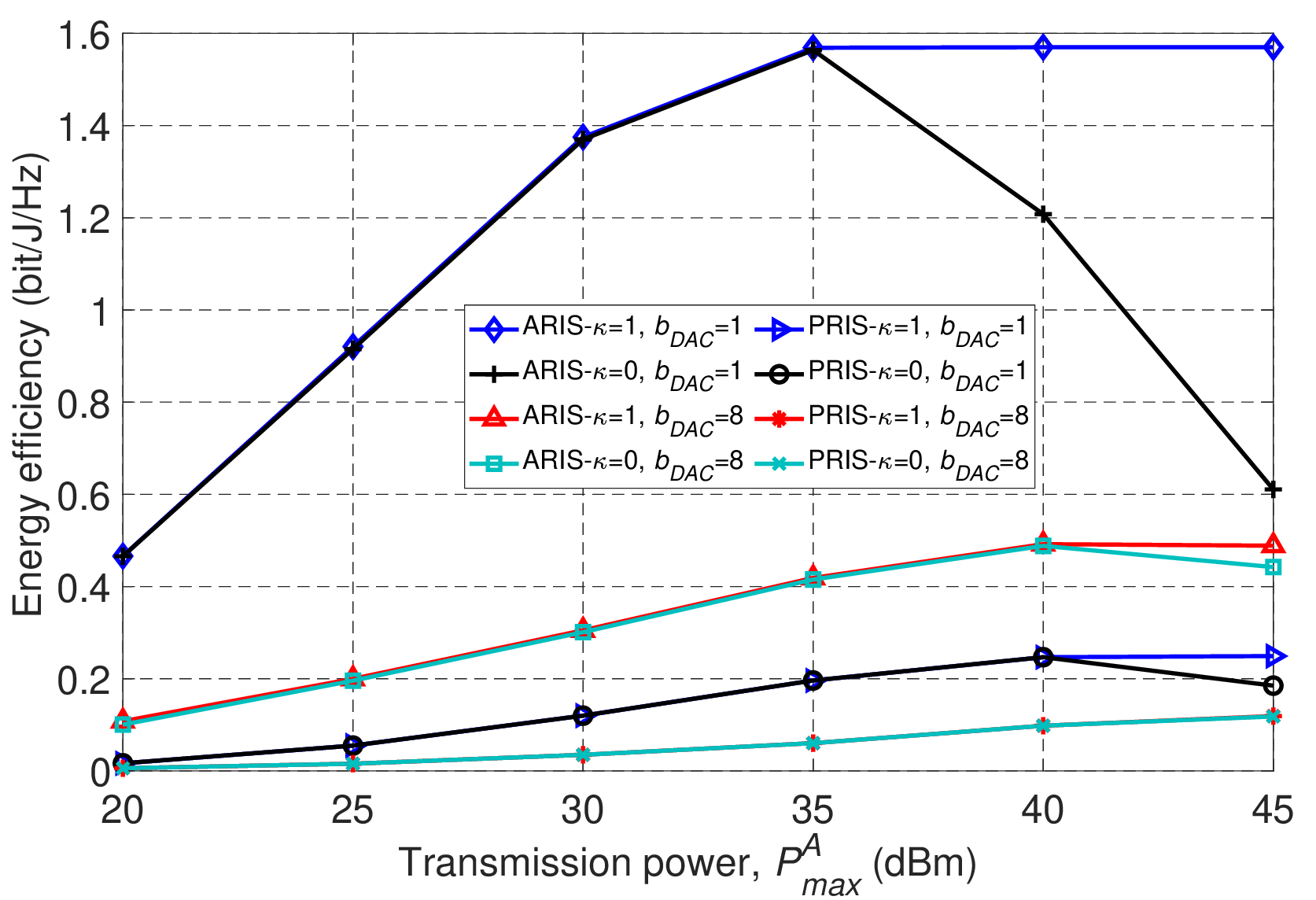} 
\caption{EE versus the maximum available transmission power. }
\label{fig_EEvsPow}
\end{figure}
\begin{figure}[!t]
	\centering
	\includegraphics[width=8.5cm,height=7cm]{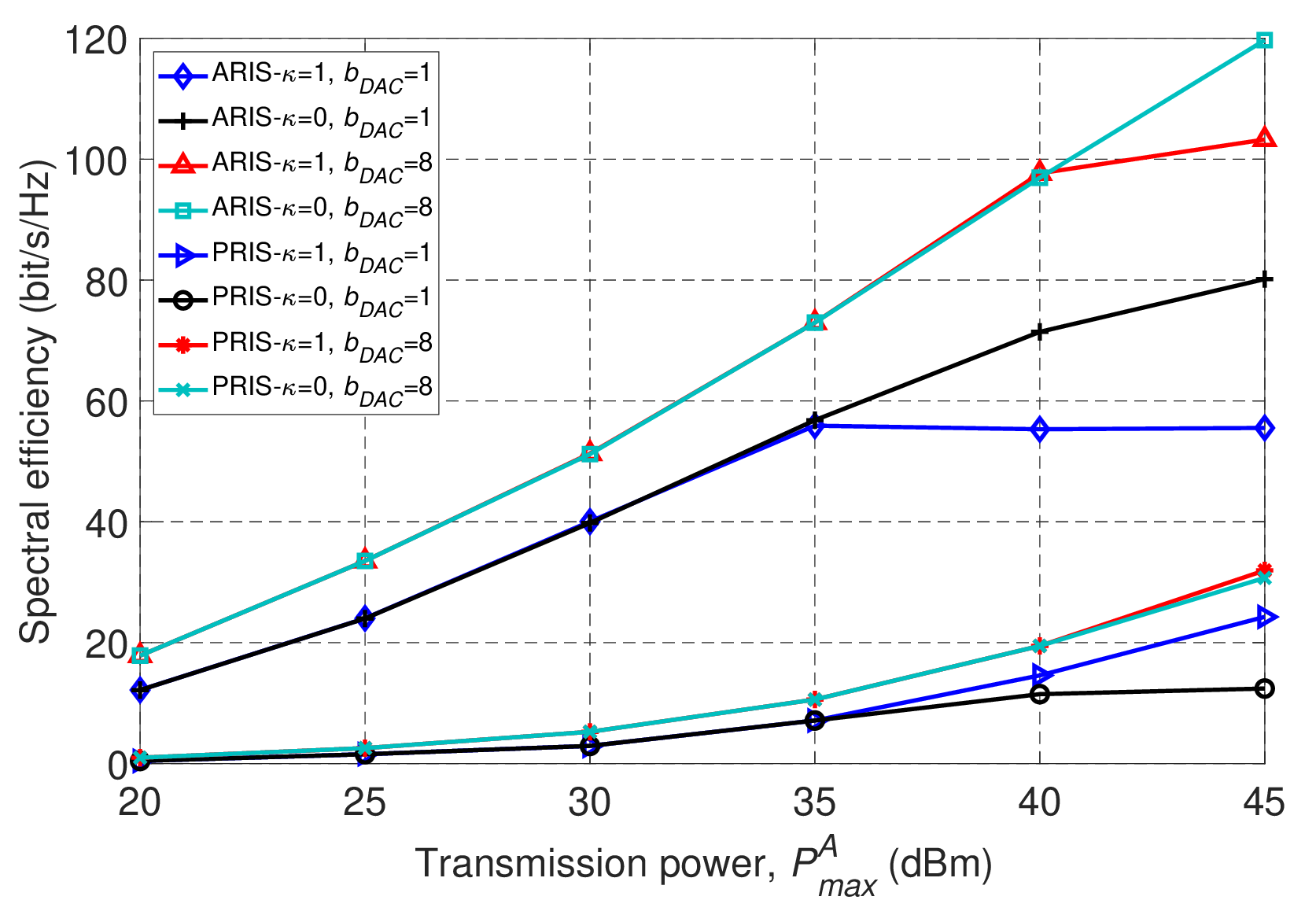} 
	\caption{SE versus the maximum available transmission power. }
	\label{fig_SEvsPow}
\end{figure}
\begin{figure*}[t]
    \centering
 \begin{minipage}{0.48\textwidth}
     \centering
	\includegraphics[width=8.5cm,height=7cm]{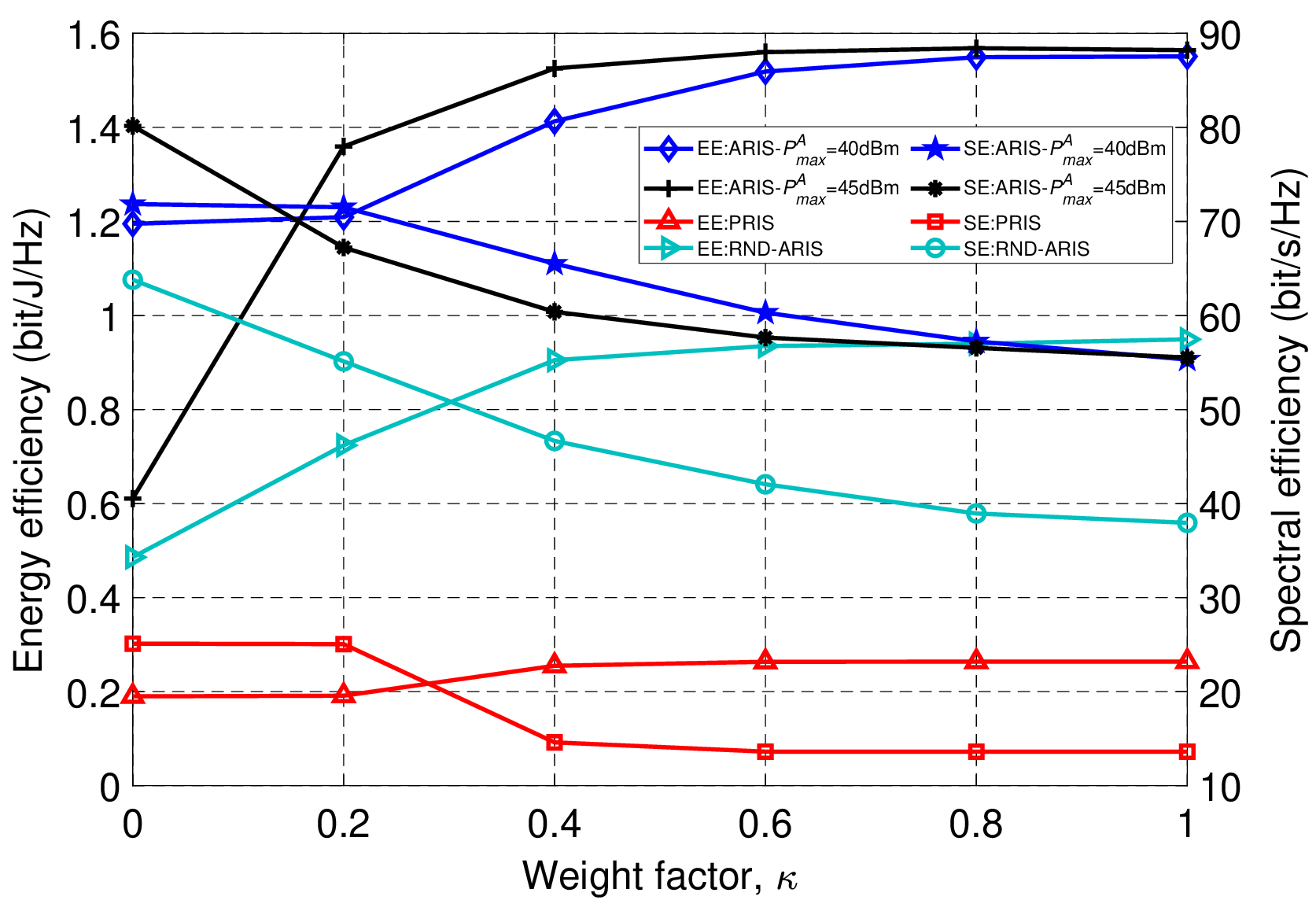} 
	\caption{EE and SE versus weight factor. }
	\label{fig_EESEvsWeight}
\end{minipage}  
      \begin{minipage}{0.48\textwidth}
        \centering
	\includegraphics[width=8.5cm,height=7cm]{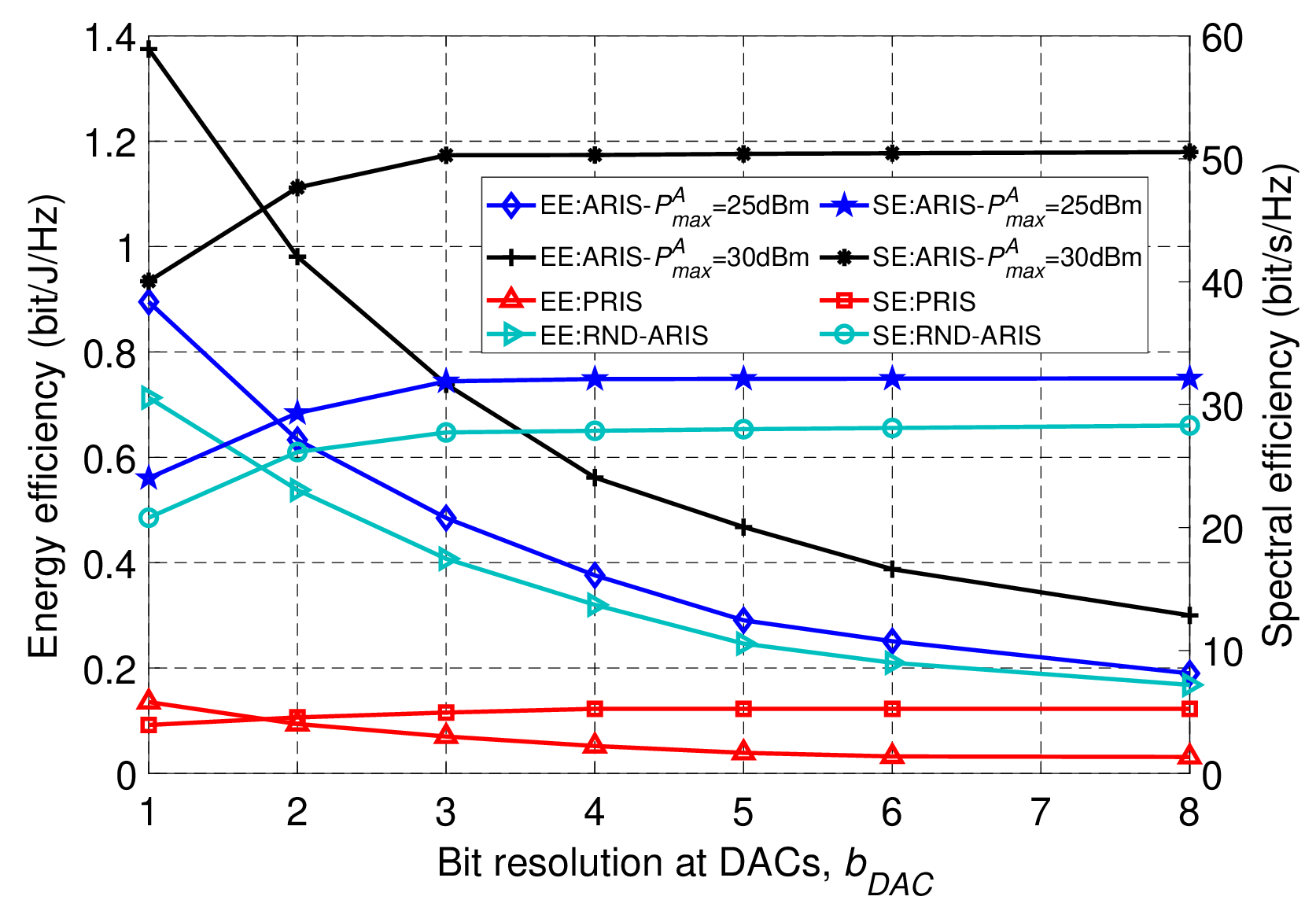} 
	\caption{EE and SE versus bit resolution at DACs. }
	\label{fig_EESEvsBit}
    \end{minipage}  
\end{figure*} 
Fig.~\ref{fig_EEvsPow} illustrates the EE as a function of the maximum available power per AP, $P_{max}^A$, under 1-bit and 8-bit DACs, considering the extremes of the SE-EE tradeoff with $\kappa=0$ and $\kappa=1$ to maximize SE and EE, respectively. We can see that the EE increases in both cases, $\kappa=0$ and $\kappa=1$, with the maximum transmission power up to $P_{max}^A=35$ dBm, while showing a significant difference at higher transmission power, where the SE-EE tradeoff becomes essential. As expected, the proposed ARIS can significantly enhance the EE over passive RIS, as active RIS is capable of directly boosting the incoming signal, which improves the achievable data rate without a significant increase in power consumption.
\begin{figure}[!t]
	\centering
	\includegraphics[width=8.5cm,height=7cm]{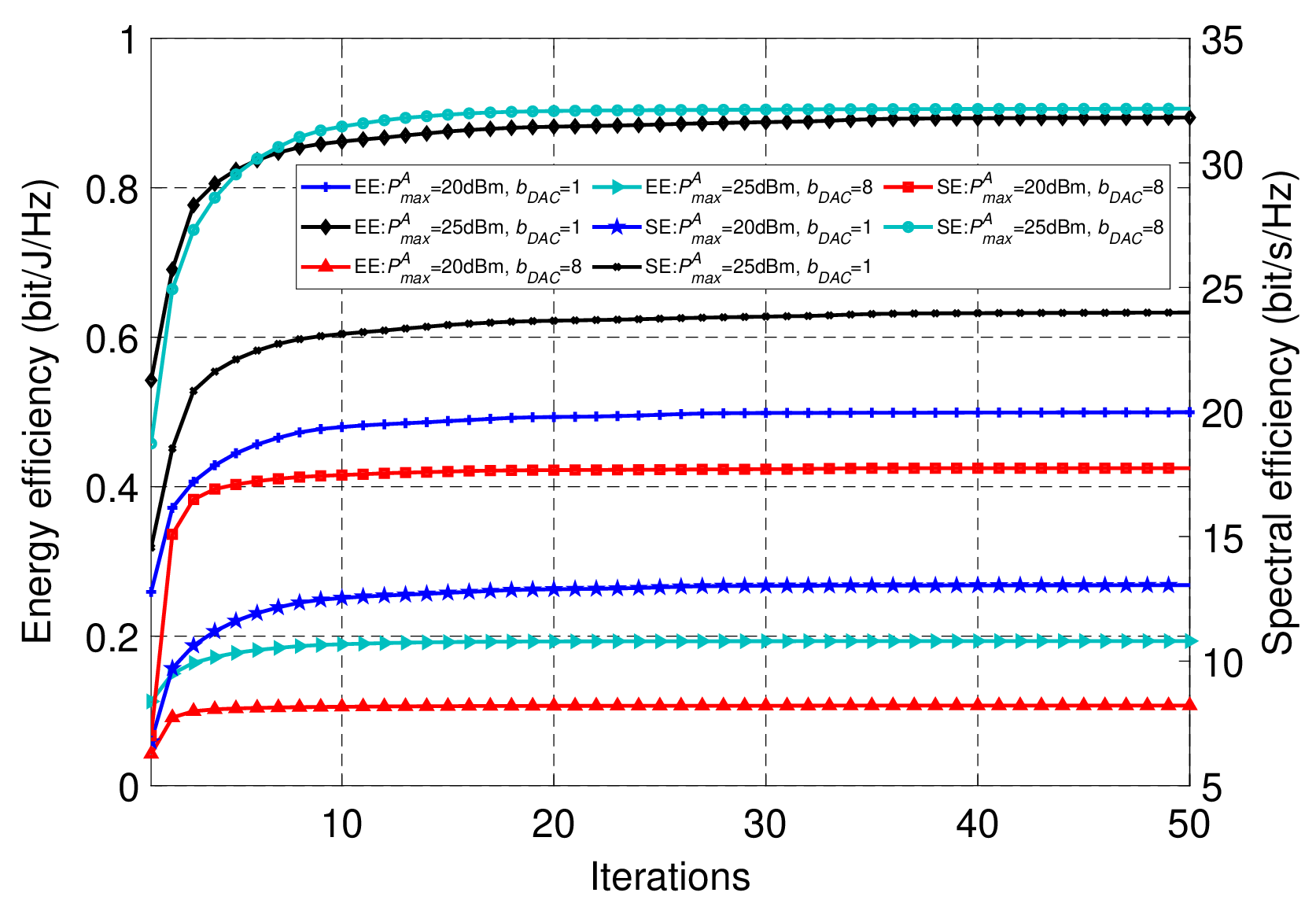} 
	\caption{Convergence behavior of the proposed algorithm. }
	\label{fig_Convergence}
\end{figure}
\begin{figure}[!t]
	\centering
	\includegraphics[width=8.5cm,height=7cm]{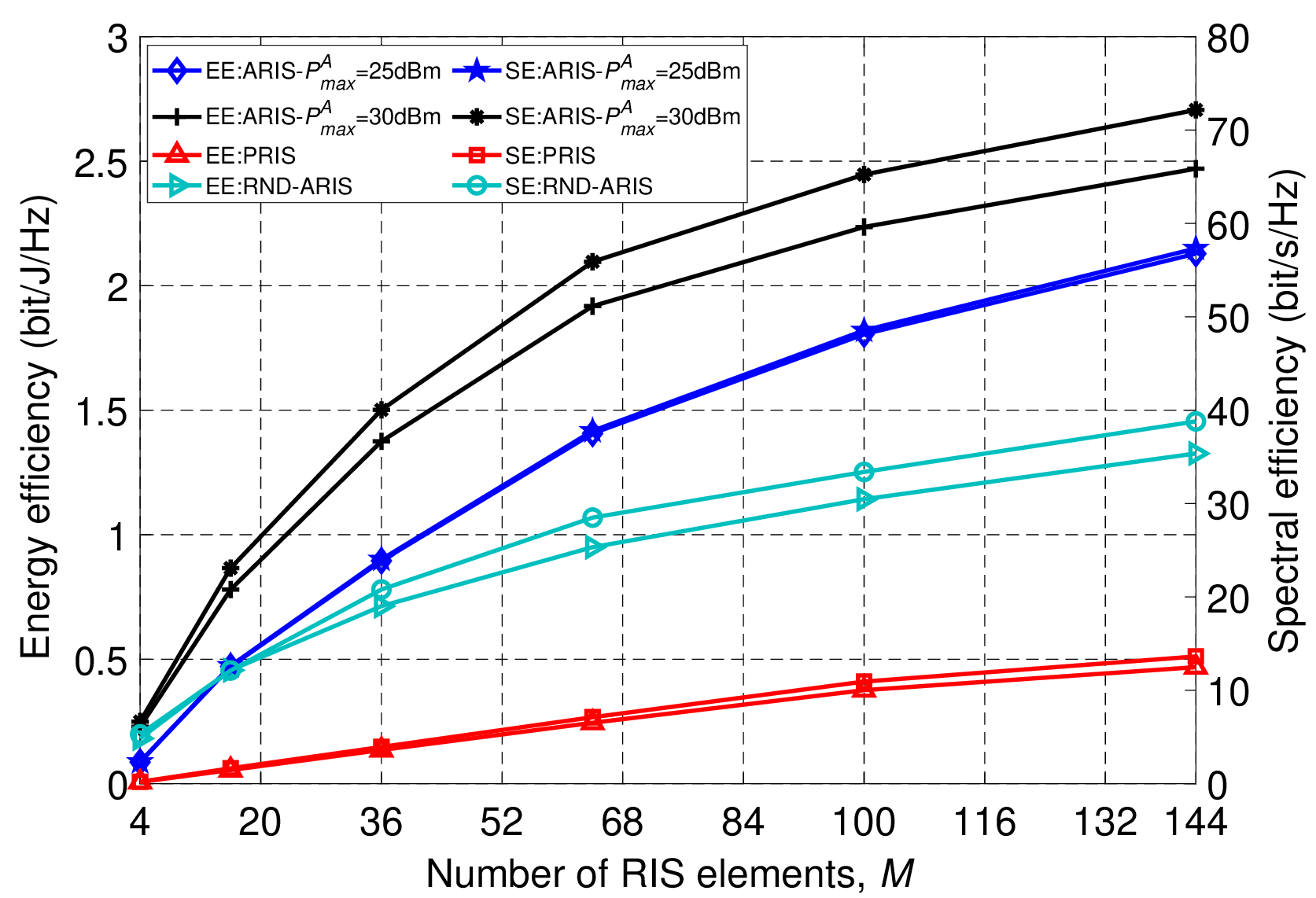} 
	\caption{EE and SE versus number of RIS elements. }
	\label{fig_EESEvsRis}
\end{figure}

Fig.~\ref{fig_SEvsPow} shows the SE as a function of the maximum available power per AP, $P_{max}^A$. We observe that the SE rises in both scenarios, $\kappa=0$ and $\kappa=1$, as the maximum transmission power increases up to $P_{max}^A=35$ dBm, while reaching a plateau for higher transmission power in the case of $\kappa=1$. From both Fig.~\ref{fig_EEvsPow} and Fig.~\ref{fig_SEvsPow}, we can see that setting $\kappa=0$ focuses on maximizing the SE without considering the power consumption, leading to a significant reduction in EE. Conversely, setting $\kappa=1$ prioritizes maximizing EE, causing the SE values-and consequently the EE-to plateau at a certain level, which does not increase even with higher transmission power.
Regarding the effect of low-resolution DACs, we observe that increasing the bits at the DACs results in a moderate increase in SE for transmission power up to $P_{max}^A=35$ dBm, which also shifts the plateau point to higher transmission power when $\kappa=1$ is used. However, higher bits at the DACs lead to a substantial reduction in EE across all transmission power ranges, demonstrating the benefits and trade-offs achievable through the use of low-resolution DACs.

Fig.~\ref{fig_EESEvsWeight} illustrates the SE and EE based on the weight factor, $\kappa$, to investigate the optimal SE-EE tradeoff. The baseline methods of PRIS and RND-ARIS are simulated with a $P_{max}^A=45$ dBm. As expected, EE increases with the value of $\kappa$, while SE decreases, demonstrating the different SE-EE tradeoffs. Moreover, we observe that the proposed scheme ARIS outperforms the baseline method RND-RIS in both SE and EE, highlighting the essential need to optimize the reflection coefficient matrices of the active RIS. Additionally, regarding the best SE-EE tradeoffs, setting $\kappa=0.4$ for $P_{max}^A=40$ dBm achieves 91\% of the maximum possible EE while reaching 91\% of the maximum SE. For $P_{max}^A=45$ dBm, using $\kappa=0.2$ achieves 87\% of the maximum possible EE while reaching 84\% of the maximum SE.

\begin{figure*}
 \begin{minipage}{0.48\textwidth}
   \centering
	\includegraphics[width=8.5cm,height=7cm]{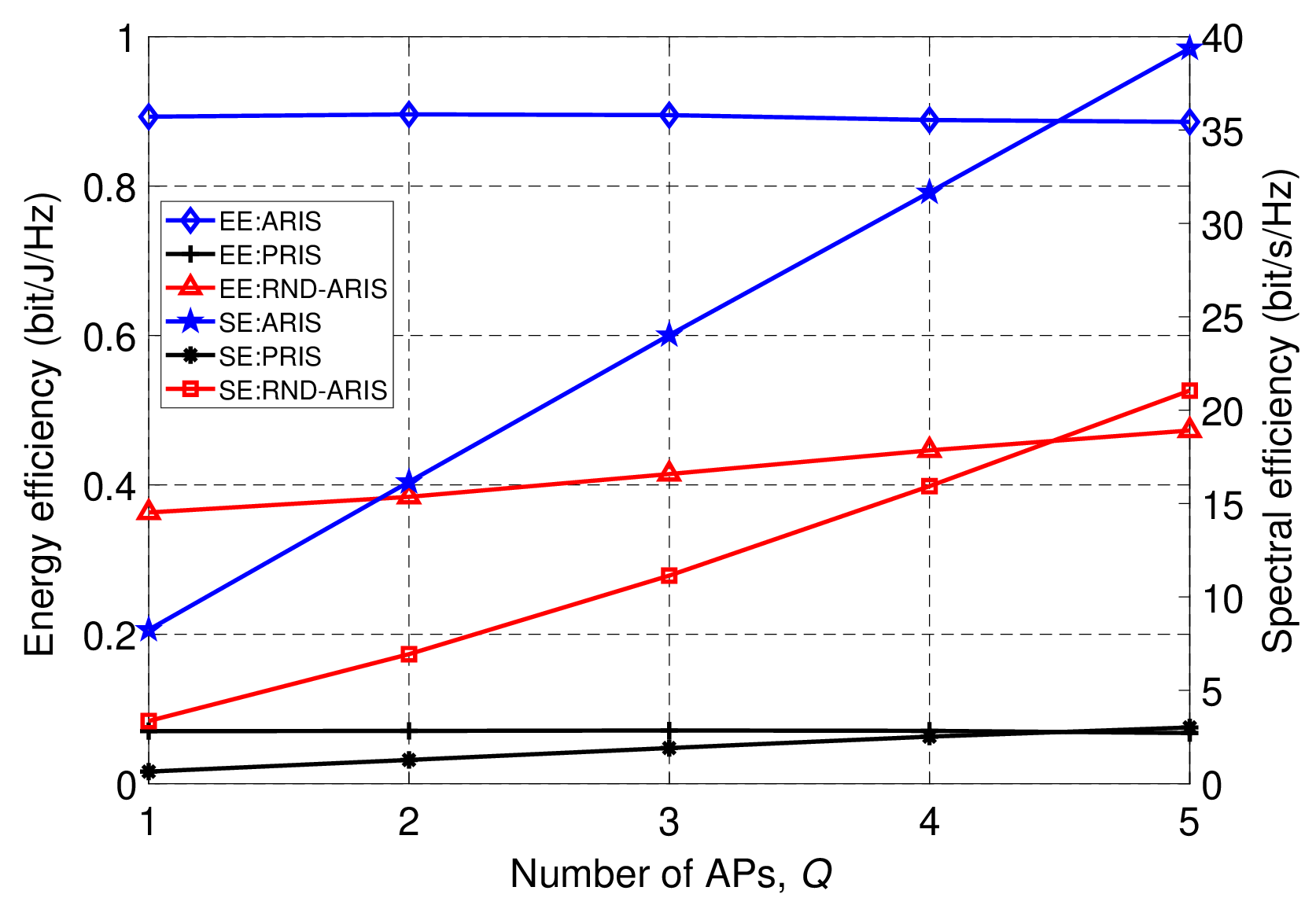} 
	\caption{EE and SE versus number of APs, $Q$. }
	\label{fig_EESEvsAP}
\end{minipage}  
 \begin{minipage}{0.48\textwidth}
     \centering
	\includegraphics[width=8.5cm,height=7cm]{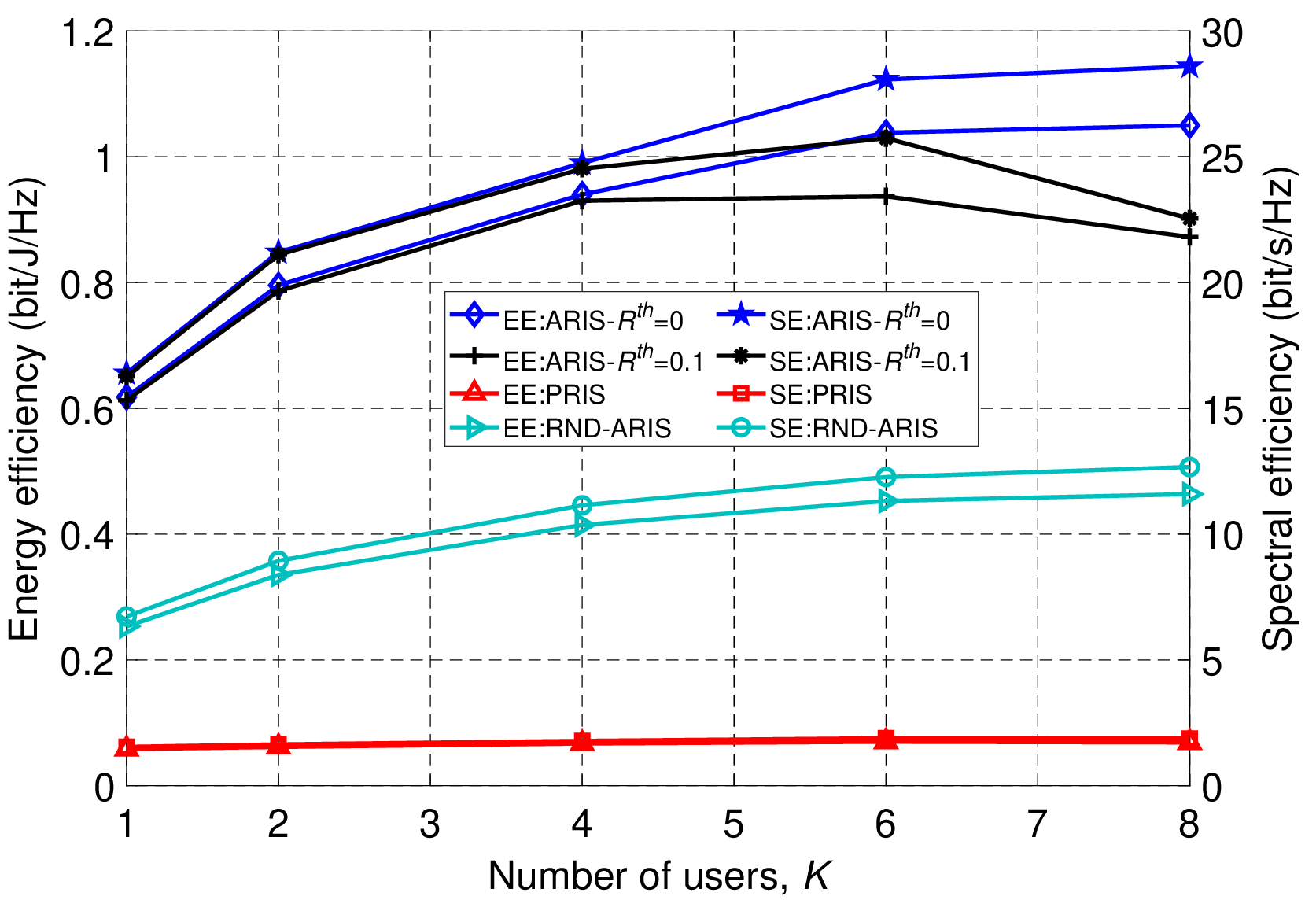} 
	\caption{EE and SE versus number of users. }
	\label{fig_EESEvsUsers}
    \end{minipage}  
\end{figure*} 

\begin{figure}
    \centering
	\includegraphics[width=8.5cm,height=7cm]{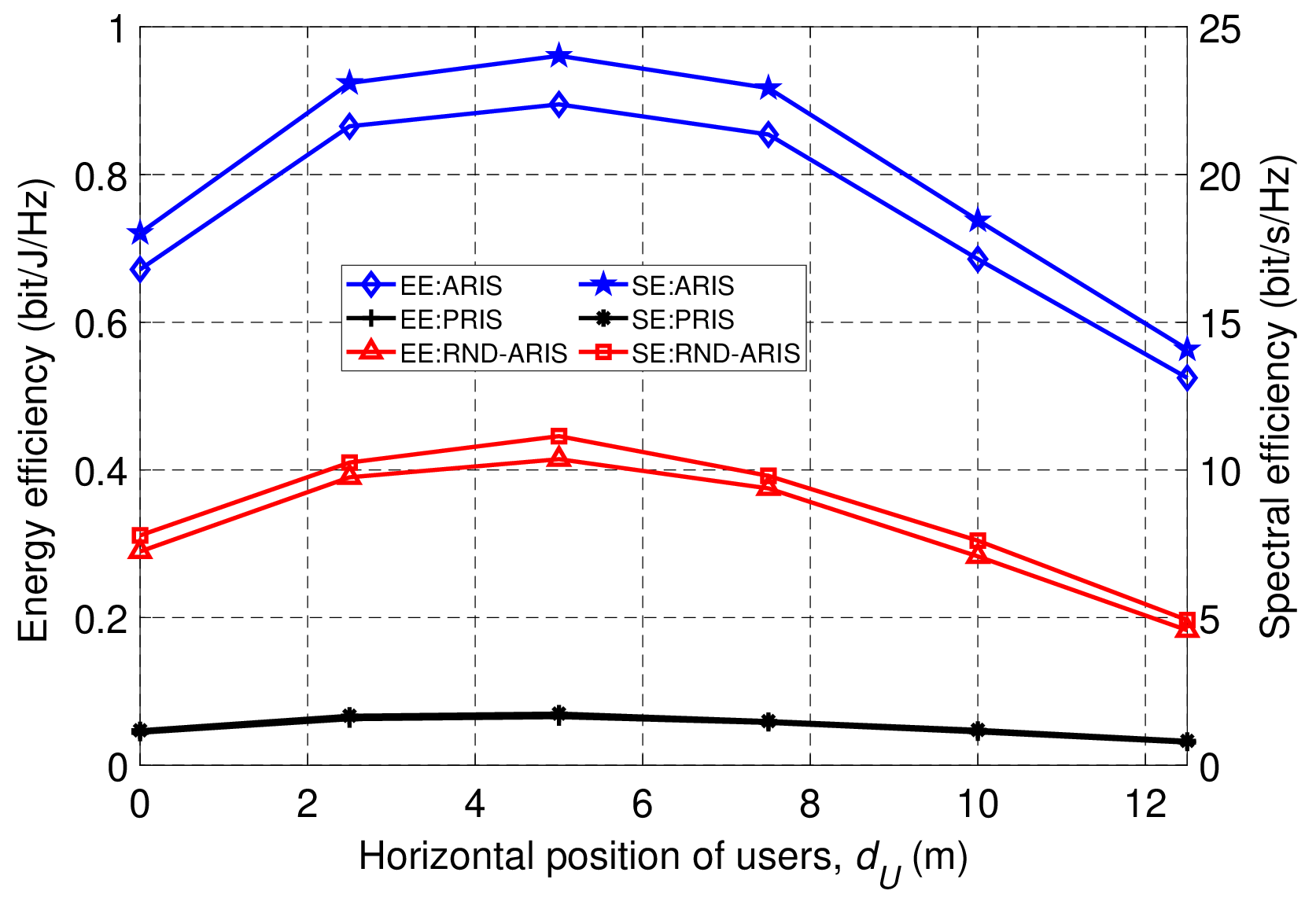} 
	\caption{EE and SE versus the horizontal position of users. }
	\label{fig_EESEvsDist}
\end{figure}
Fig. \ref{fig_EESEvsBit} presents the EE and SE according to the bit resolution, $b_{DAC}$, where the baseline methods of PRIS and RND-ARIS are simulated with $P_{max}^A=30$ dBm. Similar to Fig.~\ref{fig_EEvsPow} and Fig. \ref{fig_SEvsPow}, we observe that an increase in bit resolution leads to a moderate improvement in SE but a significant reduction in EE. Specifically, we note that beyond 4 bits at the DACs, there is negligible improvement in SE but a substantial reduction in EE. A suitable trade-off focusing on SE can be observed at $b_{DAC}=2$, where 70\% of the maximum EE is achieved while reaching 91\% to 94\% of the maximum SE. From Fig.~\ref{fig_EESEvsBit}, we conclude that using 1 or 2-bit DACs will not significantly impact the SE of the system while allowing a drastic reduction in power consumption and deployment costs.

Fig. \ref{fig_Convergence} illustrates the convergence behavior of the proposed ARIS algorithm in terms of EE and SE for different values of $P_{max}^A$ and $b_{DAC}$, with $\kappa=1$. We observe that SE and EE consistently increase with the number of iterations, achieving stable values within 10 to 25 iterations across various parameters. In particular, when $b_{DAC}=1$, 95\% of the maximum SE and EE values are achieved at around 10 iterations, while for $b_{DAC}=8$, 95\% of the maximum SE and EE values are reached at around 6 iterations.

Fig. \ref{fig_EESEvsRis} shows the EE and SE as a function of the number of reflecting elements in the RIS, $M$, with the PRIS and RND-ARIS baseline methods simulated at $P_{max}^A=30$ dBm. We observe that both EE and SE consistently increase with the number of RIS elements for all cases of passive and active RISs. This improvement is due to the additional degrees of freedom provided by the RIS elements, which enhance SE with a relatively low increase in power consumption. Moreover, we see that the proposed ARIS can achieve better SE and EE than the baseline methods of PRIS and RND-RIS across all ranges of RIS elements. The results highlight the significant role of RIS in THz systems, demonstrating its potential to achieve higher EE and SE.

Fig. \ref{fig_EESEvsAP} illustrates the EE and SE in relation to the number of APs, $Q$, considering $P_{max}^A=25$ dBm. We observe that SE significantly increases with the number of deployed APs in the system, while there are marginal changes in EE with an increasing number of APs. Increasing the number of distributed APs enhances the degrees of freedom to improve SE and EE by optimizing the precoding vectors and transmission power of each AP based on proximity and channel conditions to the RISs and UEs. This demonstrates the benefits of deploying distributed APs for RIS-assisted wideband THz systems.

Fig. \ref{fig_EESEvsUsers} shows the EE and SE as a function of the number of users, considering $P_{max}^A=25$ dBm, $\kappa=1$, and minimum rates of $R^{th}=0.1$ bit/s/Hz and $R^{th}=0$ bit/s/Hz. We observe that for the proposed ARIS, both EE and SE improve as the number of users increases up to six. However, for more than six users, SE and EE decrease in the case of $R^{th}=0.1$ bit/s/Hz because the rate constraint must be satisfied for each user on each subcarrier, leading to higher required transmission power. Conversely, without the rate constraint ($R^{th}=0$ bit/s/Hz), SE and EE consistently increase beyond six users because the optimization can focus on increasing the data rate for users with better channel conditions. 
Finally, Fig.~\ref{fig_EESEvsDist} illustrates the EE and SE in relation to the horizontal position of the users, $d_U$, considering $P_{max}^A=25$ dBm, while keeping the locations of APs and RISs fixed as specified at the beginning of Section IV. Note that the value of $d_U$ determines the users' distance along the x-axis relative to the deployed RISs. Specifically, $d_U=5$ represents the scenario where the users are closest to the RISs, while $d_U=0$ and $d_U=12$ represent the cases where the users are farthest from the RISs. As expected, as the distance from the users to RISs increases, both SE and consequently EE decrease due to the effect of path loss.

Based on the presented results, several key conclusions can be drawn regarding the EE and SE trade-offs in RIS-assisted wideband THz systems. First, the proposed ARIS approach consistently outperforms passive RIS and baseline RND-ARIS in both SE and EE, emphasizing the importance of optimizing the reflection coefficients. Our study shows that increasing transmission power enhances both EE and SE up to a point, beyond which SE saturates while EE declines—highlighting the importance of balancing performance and power usage. Moreover, low-resolution DACs, particularly $1-$ or $2$-bit, offer a promising trade-off by significantly reducing energy consumption and hardware complexity while maintaining high SE. The weight factor $\kappa$ plays a crucial role in navigating the SE-EE tradeoff, with intermediate values (e.g., $\kappa=0.2$–0.4) providing near-optimal performance in both metrics. Additionally, increasing the number of RIS elements and APs generally improves SE and EE due to greater channel manipulation capabilities and spatial diversity. The proposed algorithm shows rapid convergence within a small number of iterations, confirming its practical efficiency. Finally, the performance is sensitive to user positioning and rate constraints, indicating that intelligent user association and placement can further enhance system performance. Overall, these results validate the effectiveness of ARIS and the proposed optimization framework in enabling scalable, energy- and spectrum-efficient THz networks.

\section{Conclusion}

In this contribution, we investigated the SE-EE tradeoff in a multi-active RIS-assisted wideband THz CF-mMIMO system with low-resolution DACs. Leveraging quadratic transformation and SCA methods, we proposed a solution to jointly optimize the precoding vectors at the APs and the reflection coefficient matrices of the active RIS to achieve the best SE-EE tradeoff, subject to the minimum rate for the users, a maximum transmission power at the APs, and a maximum amplification power of the active RISs. Simulation results demonstrated that the proposed active RIS-aided CF system effectively mitigated propagation loss and limited scattering in THz communication. It achieved superior EE and SE performance compared to conventional passive RIS and random active RIS, under practical power consumption models and various system parameters. Additionally, low-resolution DACs were shown to be crucial for significantly enhancing EE while maintaining acceptable communication performance.

\vspace{12pt}

\end{document}